\documentclass[9.5pt,journal,final,finalsubmission,twocolumn]{IEEEtran}
\usepackage{url}
\usepackage{microtype,times}
\usepackage[cmex10]{amsmath}
\usepackage{cite,array,eqparbox}

\usepackage[overload]{textcase}
\usepackage{anyfontsize}
\usepackage{amsmath,amssymb,amsthm}
\usepackage{tikz}
\usetikzlibrary{trees}
\usepackage{rotating}
\usepackage{graphicx,graphics,epsfig,float,verbatim,array,bm,multirow,color}
\usepackage[left=1.5cm,top=1.5cm,bottom=1.5cm,right=1.5cm]{geometry}
\newcommand{\R}{{\mathbb R}}

\newcommand {\sr}{\stackrel}

\newcommand {\Real}{\mathbb{R}}

\newcommand {\td}{\tilde}

\newcommand {\ra}{\rightarrow}
\newcommand {\Ra}{\Rightarrow}

\newcommand {\vep}{\varepsilon}

\newcommand {\del}{\partial}
\newcommand {\Ld}{\Lambda}
\newcommand {\ld}{\lambda}

\newcommand {\al}{\alpha}

\newcommand {\bfr}{\begin{flushright}}
\newcommand {\efr}{\end{flushright}}
\newcommand {\bfl}{\begin{flushleft}}
\newcommand {\efl}{\end{flushleft}}

\newcommand {\nn} {\nonumber}

\newcommand {\txt}{\textrm}
\newcommand {\bd}{\begin{document}}
\newcommand {\ed}{\end{document}}

\newcommand {\be}{\begin{equation}}
\newcommand {\ee}{\end{equation}}
\newcommand {\bea}{\begin{eqnarray}}
\newcommand {\eea}{\end{eqnarray}}
\newcommand {\ba}{\begin{aligned}}
\newcommand {\ea}{\end{aligned}}
\newcommand {\bit}{\begin{itemize}}
\newcommand {\eit}{\end{itemize}}
\newcommand {\ul}{\underline}
\newcommand {\txtc}{\textcolor}

\newcommand {\ad}{\textrm{ad}}
\newcommand {\Ad}{\textrm{Ad}}
\def\A{{\cal A}}\def\B{{\cal B}}\def\C{{\cal C}}\def\D{{\cal D}}\def\E{{\cal E}}\def\F{{\cal F}}\def\G{{\cal G}}\def\H{{\cal H}}\def\I{{\cal I}}
\def\J{{\cal J}}\def\K{{\cal K}}\def\L{{\cal L}}\def\M{{\cal M}}\def\N{{\cal N}}\def\O{{\cal O}}\def\P{{\cal P}}\def\Q{{\cal Q}}\def\R{{\cal R}}
\def\S{{\cal S}}\def\T{{\cal T}}\def\U{{\cal U}}\def\V{{\cal V}}\def\W{{\cal W}}\def\X{{\cal X}}\def\Y{{\cal Y}}\def\Z{{\cal Z}}

\def\xbar{\bar{x}}\def\ybar{\bar{y}}\def\zbar{\bar{z}}\def\kbar{\bar{k}}\def\pbar{\bar{p}}

\newtheorem{thm}{Theorem}[section]
\newtheorem{lmm}{Lemma}[section]
\newtheorem{dfn}{Definition}[section]
\newtheorem{crl}{Corollary}[section]
\newtheorem{prp}{Proposition}[section]
\newtheorem{prb}{Problem}[section]
\newtheorem{alg}{Algorithm}[section]
\newtheorem{rmk}{Remark}[section]
\newtheorem{assumption}{Assumption}[section]
\theoremstyle{definition}\newtheorem{example}{Example}[section]
\theoremstyle{definition}\newtheorem*{rmks}{Remarks}
\newcommand*{\Cdot}[1][1.25]{%
  \mathpalette{\CdotAux{#1}}\cdot%
}
\newdimen\CdotAxis
\newcommand*{\CdotAux}[3]{%
  {%
    \settoheight\CdotAxis{$#2\vcenter{}$}%
    \sbox0{%
      \raisebox\CdotAxis{%
        \scalebox{#1}{%
          \raisebox{-\CdotAxis}{%
            $\mathsurround=0pt #2#3$%
          }%
        }%
      }%
    }%
    \dp0=0pt %
    \sbox2{$#2\bullet$}%
    \ifdim\ht2<\ht0 %
      \ht0=\ht2 %
    \fi
    \sbox2{$\mathsurround=0pt #2#3$}%
    \hbox to \wd2{\hss\usebox{0}\hss}%
  }%
}

\begin{document}
\title{On the Sequential Test and Distributed Detection}

\author{
\author{Earnest~Akofor
\thanks{E. Akofor is with the Department of Mathematics and Computer Science, Faculty of Science, University of Bamenda, PO Box 39 Bambili, NW Region, Cameroon, e-mail: eakofor@gmail.com.}%
}
}

%



\maketitle

\begin{abstract}
We present a simple definition of stopping time and its role in the formulation of sequential tests for both centralized and distributed detection, providing a straightforward procedure for obtaining optimal decision rules. Upper bounds for optimal stopping time are derived and numerically shown to possess certain qualitative features expected of the optimal stopping time. The results are extended to any distributed detection network in the form of an acyclic directed graph.
\end{abstract}

\begin{keywords}
Network decision rules, Centralized sequential detection, Distributed sequential detection, Stopping time bounds, Acyclic directed graph networks.\\
\indent\emph{\textbf{2020 MSC---}} 62L10, 62L15, 94A13, 62C10, 60G40.
\end{keywords}

\let\thefootnote\relax\footnotetext{This work began in 2014 while the author was still with the Department of EECS, Syracuse University, Syracuse, NY 13244, USA and was partly supported by Air Force Office of Scientific Research under Award FA9550-10-1-0458, by Army Research Office under Award W911NF-12-1-0383, and by National Science Foundation under Award CCF1218289.}

\tableofcontents

\section{Introduction}
The work done in this paper was hinted at in the \emph{conclusion} of the dissertation \cite[Section 7.3]{akofor2016}, under the title ``\emph{Comments on sequential detection}''.

It is a well known fact in detection theory that error probability decreases as observation sample size increases. In some applications, instead of trying to move the probability of error arbitrarily close to 0, we may be willing to tolerate a certain level of nonzero error probability $\vep$. In this case, we can shift our interest in the direction of finding the smallest possible number of observation samples necessary to reach a level of accuracy such that error probability does not exceed $\vep$.

A classical testing procedure for the situation described above is Wald's sequential test which was introduced in \cite{wald-45}. For us, \emph{sequential detection} refers to detection with \emph{sampling time} as the performance metric. The phrases ``\emph{sample size}'', ``\emph{sample number}'', and ``\emph{stopping time}'' will be used interchangeably, and it will sometimes be assumed that these phrases are preceded by the word(s) ``\emph{optimal}'' or ``\emph{expected}''. Sequential detection is dynamic in the sense that the decision rule is a random process. The sensor persistently obtains and processes an increasing sample of several (one-dimensional) observations until a certain stopping criterion is met. Each observation is taken and processed together with previous processing results and previous observations, assuming the processor has memory, before the next observation is taken. This is unlike in static detection where the sensor takes only one observation sample of fixed size.

Detail on the formulation and performance analysis of the centralized sequential test is found in \cite{wald-45}, \cite{wald-47-book}, and other standard references on the sequential test, such as \cite{TtkvEtal2014}. Likewise, detailed formulation and analysis of the distributed sequential test based on Wald's original ideas exist in the references given below. Therefore, we will not be concerned with the familiar aspects of the sequential test. Instead, our main gaol is to provide a simple alternative description of the transitional step between the centralized and distributed sequential tests. It is our hope that the relatively simple procedure used to derive the sequential tests here can serve as a convenient tool for sequential testing in more involved settings, including and beyond detection over arbitrary networks in the form of acyclic directed graphs.

Distributed sequential detection without fusion was studied in \cite{teneketzis-ho-87}, where dynamic programming techniques were proposed for computing thresholds. The same non-fusion setup, based on an earlier version of \cite{teneketzis-ho-87}, was considered in \cite{lavigna-makowski-baras-86} under continuous time, and some advantages over the discrete time version were noted. A sequential distributed detection problem with a parallel fusion architecture was later studied in \cite{hashemi-rhodes-89}, wherein it was pointed out that an optimal decision strategy required the thresholds of each sensor to depend on its past decisions. A comment on the above reference is given in \cite{Veeravalli-92-comments}. A review of earlier results and a problem setting with a more general fusion architecture can be found in \cite{Veeravalli-92-thesis}, while fusion architectures involving feedback were studied in \cite{Veeravalli-92-thesis, veeravalli-basar-poor-93, veeravalli-99}. Some asymptotic aspects of the distributed sequential test with a parallel fusion architecture have been considered in \cite{mei-2008}. In \cite{nayyar-teneketzis-2009}, distributed sequential detection is considered with a tandem fusion architecture, where it is shown that three-threshold rules may not be optimal for certain finite horizon problems.

In most existing work, dynamic programming techniques (which use Bayes' rule to update prior information at each observation/optimization instant) are used to show that the optimal stopping rule is such that the dynamic optimization process, hence the observation process, stops whenever a desired level of accuracy is reached. Because of the form of the stopping rule, the expected stopping time is primarily a property of the observations. These dynamic optimization approaches are in line with the usual estimation of the expected sample number using Wald's identities derived in \cite{wald-44}, for the expected number of \emph{independent} variables whose values lie outside of a fixed (time-independent) value range, or using the generalizations of such identities in \cite{chernoff-59}.

Our approach differs in the following way. We consider a sequence of \emph{ternary}, or \emph{three-value}, decisions for which \emph{sampling time} is explicitly introduced as a random variable whose probability distribution is simply given by the probability of one of the three possible decision values. A desired level of accuracy is maintained throughout a possibly \emph{infinite} decision process in which the \emph{instantaneous} decision whether to continue sampling or to stop is made in such a way that the expected number of observations is minimized. Thus, the expected stopping time in our approach is primarily a property of the decision strategy. This interpretation of stopping time greatly simplifies our analysis in comparison with other formulations of the sequential detection problem. Moreover, our approach and results can be extended to more complex fusion architectures with relative ease.

Using a detection network optimization method established in \cite{IT-paper,akofor2016,AkoChen013I,AkoChen013II,ZhuAkoChen013}, we obtain optimal decision rules for the centralized and distributed tests. To avoid exact numerical computation involving an infinite number of thresholds, we derive upper bounds (for stopping time) based on \emph{sample-by-sample} processing versions of the sequential test in which the current decision is based on the current observation and previous decisions but ignores previous observations. These upper bounds of stopping time, which are themselves performance metrics for \emph{partial memory} versions of the \emph{full memory} sequential tests, are observed to provide useful insight on the qualitative properties of the optimal stopping time. Numerical results show that the behaviors of these bounds agree with what is intuitively expected of the dependence of stopping time on parameters such as observation quality and the desired levels of accuracy. The above findings are shown to readily generalize to an arbitrary detection network in the form of an acyclic directed graph. This extension is remarkable since any given distributed detection network, no matter how elaborate, can be approximated by an acyclic directed graph. For more on related work, see \cite{ZhangEtal2020} and the references therein.

The rest of the work is organized as follows. We begin with a description of the decision procedure for centralized and distributed detection networks in Section \ref{dp-section}. This is followed by a discussion of the centralized sequential test in Section \ref{cst-section}, along with some numerical results for the centralized version of the distributed sequential test to be described later. The distributed sequential test for a simple two-sensor tandem network is then presented in Section \ref{dst-section}. In Section \ref{generalize}, the main results of the previous sections are generalized to distributed detection over any sensor network in the form of an acyclic directed graph. We conclude our discussion in Section \ref{cnl-section}. The proofs of some two main theorems are given as appendices.

\emph{Notation:} We label sensors using upper case letters $X$, $Y$, $Z$, and so on. Random variables and their values are both denoted by lower case letters $x$, $y$, $z$, $u$, $v$, $w$, etc. Likewise, we do not distinguish between summation and integration symbols, i.e., if $u$ is discrete and $x$ is continuous, we write $\sum_{u,x}=\sum_u\sum_x$, where $\sum_u$ denotes summation over $u$, and $\sum_x$ denotes integration over $x$. Similarly, if $u$, $v$ are discrete and $x$, $y$ are continuous, we write $\delta_{(u,x)(v,y)}=\delta_{uv}\delta_{xy}$, where $\delta_{uv}$ is the Kronecker delta, while $\delta_{xy}$ is the Dirac delta.  

\section{General Decision Procedure}\label{dp-section}
The basic decision theory has been developed in \cite[Section II]{IT-paper} and in \cite{akofor2016}. Nevertheless, we will only borrow one major result from the above references, while providing all intermediate steps which are relevant for our discussion.

For concreteness, we first describe, in Section \ref{single-sensor}, the \emph{simplest} optimal decision rule for a single sensor in isolation. Such a decision rule of course also applies to any centralized system of sensors, i.e., a system of sensors for which all observations are available in the same location. Next, we generalize the discussion of Section \ref{single-sensor} to obtain decision rules for a distributed sensor network in Section \ref{sensor-network}.

Note that in Section \ref{single-sensor}, the basic decision rule considered assumes the sensor makes a \emph{one-step} decision based on the observation only. The case of a single sensor making multiple decisions (e.g., a sequence of decisions with memory) is a special case of the distributed network rules of Section \ref{sensor-network}. This is because a single sensor making $n$ decisions can always be equivalently viewed as a distributed system of $n$ sensors each making a single decision.

Finally, note that the affine objective functions we will consider in this section, although relevant for \emph{static detection} in their own right, are only for illustration of the general optimization procedure. The objective functions for \emph{sequential detection} we shall encounter in Sections \ref{cst-section}, \ref{dst-section}, and \ref{generalize} are non-affine convex functions. However, by a result of \cite{IT-paper}, exactly the same optimization technique applies.

\subsection{Single sensor rules (for one-step decisions)}\label{single-sensor}
Suppose a \emph{discrete} random signal $s\in \S$ is observed by an isolated sensor $X$ as
\bea
\label{signal-transform1}x=s+b~\in~\X,
\eea
where $b\in\B$ is a continuous random parameter whose distribution is continuous and known, and $s,b$ are \emph{statistically independent} of one another. Let the signal alphabet be given by $\S=\{\mu_0,\mu_1,...,\mu_{M-1}\}$. Then we have a set of $M$ \emph{hypotheses}
\bea
H_i:~s=\mu_i,~~\txt{i.e.,}~~x=\mu_i+b,~~~~~~~~i=0,1,...,M-1,\nn
\eea
each of which represents a possible value of the signal $s$.
For the special case where $M=2$ and $\S=\{0,1\}$, the hypothesis $H_0$ denoting ``target absent'' is called the \emph{null} hypothesis, while the hypothesis $H_1$ denoting ``target present'' is called the \emph{alternative}  hypothesis.

We denote the conditional probability density function, or pdf, $p(x|H_i)=p(x|s=\mu_i)$ of $x$ under $H_i$ by $p_i(x)$. Note that $p_i(x)$ is known for each $i$ since the distribution of $b$ in (\ref{signal-transform1}) is known. A \emph{quantization rule} of the sensor $X$ is a mapping
\bea
\gamma:x\in\X~\longmapsto~ u=\gamma(x)\in \U=\{0,1,...,N-1\},\nn
\eea
where we often denote the rule $\gamma$ by its value $u$ as a \emph{variable}. When $N=M$, the quantization rule is called a \emph{decision rule}, and the \emph{decision} $u=i$ is then naturally interpreted as \emph{acceptance} of the hypothesis $H_i$. However, we will refer to the quantization rule as a decision rule even when $N\neq M$.

Our objective is to choose the rule $\gamma$ such that some function $S=S(\gamma(x))$ of $u=\gamma(x)$ is optimized. Since $u$ is a random variable, it suffices to treat $S$ as a function of the conditional distributions $p(u|x)$, for all $x\in \X$. For illustration, we consider the Bayesian objective function
\bea
\label{bayes-objective1}S = \sum_{i,j}C_{ij}p(u=i,H_j)=\sum_{i,x,j}C_{ij}p(u=i|x)p_j(x)\pi_j,
\eea
where $C_{ij}$ is a nonnegative number denoting the cost of deciding $H_i$ when/while $H_j$ is true, $p_j(x)=p(x|H_j)$, and $\pi_j=p(H_j)=p(s=\mu_j)$ is the probability that $H_j$ is true.

Since $S$ is an affine function of the conditional probabilities $p(u|x)$ and the observation $x$ is continuously distributed, by \cite[Proposition II.1]{IT-paper} the optimal decision rule is deterministic and, for each $i=0,1,...,M-1$, is given by
\bea
\label{optimal-rule1}p_{\txt{opt}}(u=i|x)=I_{R_{u=i}}(x)=\left\{
                                                            \begin{array}{ll}
                                                              1, & \hbox{if}~~x\in R_{u=i}, \\
                                                              0, & \hbox{if}~~x\not\in R_{u=i},
                                                            \end{array}
                                                          \right.
\eea
where the \emph{decision region} $R_{u=i}$ is given by

{\small
\be\ba
&\label{optimal-region1}R_{u=i}=\bigcap_{j\neq i}\left\{x:{\del S\over\del p_{\txt{opt}}(u=i|x)}-{\del S\over\del p_{\txt{opt}}(u=j|x)}<0\right\}\\
&~~~~=\bigcap_{j\neq i}\left\{x:\sum_lC_{il}p_l(x)\pi_l-\sum_lC_{jl}p_l(x)\pi_l<0\right\}.
\ea\ee
}

In terms of probabilities of the decision regions, the optimal value of $S$ is
\bea
\label{optimal-bayes-objective1}S_{\txt{opt}}=\sum_{i,j}C_{ij}~\pi_j~p_j(R_{u=i}).
\eea

\subsection{Sensor network rules}\label{sensor-network}
Now suppose we have a distributed network of $n$ sensors $\{X_k,~k=1,...,n\}$, each sensor observing the same signal $s\in \{\mu_0,\mu_1,...,\mu_{M-1}\}$. Here, the main difference with the preceding section is that each $X_k$ must now make its decision $u_k$ based not only on its own observation $x_k$, but as well on the set of decisions $\td{u}_k$ of all sensors forwarding their decisions to $X_k$, i.e., $u_k=\gamma_k(x_k,\td{u}_k)$ for some integer-valued function $\gamma_k$. Thus, $X_k$ makes an observation
\bea
\label{signal-transform2}x_k=s+b_k~\in~\X_k,~~~~b_k\in\B_k,
\eea
considers a set of hypotheses
\bea
H_i:~s=\mu_i,~~\txt{i.e.,}~~x_k=\mu_i+b_k,~~~~i=0,1,...,M_k-1,\nn
\eea
and applies a quantization rule
\bea
\gamma_k:(x_k,\td{u}_k)\in \X_k\times\td{\U}_k~\longmapsto~u_k=\gamma_k(x_k,\td{u}_k)\in \U_k,\nn
\eea
where $\U_k=\{0,1,...,N_k-1\}$, and $\td{u}_k\in \td{\U}_k$ denotes the \emph{set} of decision variables of all sensors transmitting their decisions to $X_k$. Here, for each $k$, we again assume the parameters $s$, $b_k$ in (\ref{signal-transform2}) have the same properties as before. We further assume without loss of generality that $N_k=N$ is fixed for all $k=0,1,...,n$, in which case, $\td{\U}_k=\{0,1,...,N-1\}^{I_k}$, where $I_k$ is the number of sensors transmitting their decisions to $X_k$.

Without loss of generality, we will let sensor $X_1$ serve as the fusion center for the network. Once again, our objective is to choose the \emph{decision strategy} $\{\gamma_1,\gamma_2,...,\gamma_n\}$ such that some function $S=S\big(\gamma_1(x_1,\td{u}_1),\gamma_2(x_2,\td{u}_2),...,\gamma_k(x_n,\td{u}_n)\big)$ of $u_1=\gamma_1(x_1,\td{u}_1)$, $u_2=\gamma_2(x_2,\td{u}_2)$, $\cdots$, $u_n=\gamma_n(x_n,\td{u}_n)$ is optimized. For illustration, we consider $S$ to be the Bayesian objective function at $X_1$, i.e.,
\be\ba
&\label{bayes-objective2}S = \sum_{i,j}C_{ij}p(u_1=i,H_j)=\sum_{u_1,i}C_{u_1i}p(u_1,H_i)\\
&~~~~=\sum_{u^n,x^n,i}C_{u_1i}~\prod_{k=1}^np(u_k|x_k,\td{u}_k)~p_i(x^n)\pi_i,
\ea\ee
where $u^n=u_1,...,u_n$, $x^n=x_1,...,x_n$.


Since $S$ is an affine function of the conditional probabilities $p(u_k|x_k,\td{u}_k)$ and the observations $x_k$ are continuously distributed, by Proposition II.1 of \cite{IT-paper} the optimal decision rule for each sensor $X_k$ is deterministic, and is given by
\bea
\label{optimal-rule2}p_{\txt{opt}}(u_k|x_k,\td{u}_k)=I_{R_{u_k|\td{u}_k}}(x_k)=\left\{
                                                            \begin{array}{ll}
                                                              1, & \hbox{if}~~x_k\in R_{u_k|\td{u}_k}, \\
                                                              0, & \hbox{if}~~x_k\not\in R_{u_k|\td{u}_k},
                                                            \end{array}
                                                          \right.
\eea
where $u_k=0,1,...,N_k-1$, and the \emph{decision region} $R_{u_k|\td{u}_k}$ is given by

{\footnotesize
\bea
\label{optimal-region2}R_{u_k|\td{u}_k}=\bigcap_{u\neq u_k}\left\{x_k:{\del S\over\del p_{\txt{opt}}(u_k|x_k,\td{u}_k)}-{\del S\over\del p_{\txt{opt}}(u|x_k,\td{u}_k)}<0\right\}.
\eea
}The derivative of $S$ in (\ref{optimal-region2}) can be express as

{\footnotesize
\be\ba
\label{optimal-derivative2}&{\del S\over\del p_{\txt{opt}}(u_k|x_k,\td{u}_k)}=\sum_{\{u^n,x^n\}_k,~i}C_{u_1i}~\prod_{k'\neq k}p_{\txt{opt}}(u_{k'}|x_{k'},\td{u}_{k'})~p_i(x^n)\pi_i\\
&~~~~=\sum_{\{u^n,x^n\}_k,~i}C_{u_1i}~\prod_{k'\neq k}I_{R_{u_{k'}|\td{u}_{k'}}}\left(x_{k'}\right)~p_i(x^n)\pi_i,
\ea\ee
}where $\{u^n,x^n\}_k=\{u^n,x^n\}\backslash\{u_k,x_k,\td{u}_k\}$. With \emph{conditionally independent observations}, we have $p_i(x^n)=\prod_{k=1}^np_i(x_k)$. If we further assume there are \emph{no closed processing paths} in the sensor network that can lead to \emph{overlaps} among the decision regions, then (\ref{optimal-derivative2}) can be written as

{\footnotesize
\be\ba
\label{optimal-derivative2.2}{\del S\over\del p_{\txt{opt}}(u_k|x_k,\td{u}_k)}=\sum_{\{u^n,x^n\}_k,~i}C_{u_1i}~\prod_{k'\neq k}p_i\left(R_{u_{k'}|\td{u}_{k'}}\right)~p_i(x_k)\pi_i,
\ea\ee
}in which case, the optimal value of $S$ in terms of probabilities of the decision regions is
\bea
\label{optimal-bayes-objective2}S_{\txt{opt}}=\sum_{u^n,i}C_{u_1i}~\pi_i~\prod_{k=1}^np_i\left(R_{u_k|\td{u}_k}\right).
\eea
We have thus proved the following result.
\begin{thm}\label{acyclic-prop}
Suppose we are given a network of $n$ sensors $X_1,X_2,...,X_n$ with conditionally independent and continuously distributed observations $x_1,x_2,...,x_n$. Suppose further that there are no cycles in the sensor network (i.e., the sensor network is an acyclic directed graph with the sensors as nodes and decision transmissions as arrows). Then the decision rules for detection, based on the Bayes function (\ref{bayes-objective2}), by the sensor network are given by (\ref{optimal-rule2}), (\ref{optimal-region2}), and (\ref{optimal-derivative2.2}). Moreover, the optimal value of the Bayes function is given by (\ref{optimal-bayes-objective2}).
\end{thm}

In the rest of our discussion we will consider only \emph{binary} hypothesis and \emph{ternary} decisions, i.e., we set $M=2$ and $N=3$. When the problem is sufficiently simple, like the ones we will study, the decision regions (\ref{optimal-region2}) can be completely specified in terms of a number of \emph{variable} threshold parameters that do not depend on the observations. In that case, we only need to optimize $S$ as a function of the thresholds. For both the centralized sequential test in Section \ref{cst-section} and the distributed sequential tests in Sections \ref{dst-section} and \ref{generalize}, we will carry out this optimization procedure explicitly.  

\section{The Centralized Sequential Test}\label{cst-section}
Consider an $n$-sample observation $x^n=(x_1,...,x_n)$, \emph{two} hypotheses
\be
H_j:x^n\sim p_i(x^n)=p(x^n|H_i),~~~~i=0,1,\nn
\ee
and a \emph{ternary} decision rule
\be\ba
&\gamma_n:(x^n,u^{n-1})\in\X^n\times\U^{n-1}\nn\\
&~~~~\longmapsto u_n=\gamma_n(x^n,u^{n-1})\in \U_n=\{0,c,1\}.\nn
\ea\ee
Here, $u_n=0$ is acceptance of $H_0$, $u_n=1$ is acceptance of $H_1$, and $u_n=c$ is rejection of both $H_0$ and $H_1$ in favor of continued sampling. The next sample $x_{n+1}$ is taken if and only if $\gamma_t(x^t,u^{t-1})=c$ for all $t\leq n$. This implies the sample size $n=g_n(u^{n-1})$ is a random variable whose distribution $p(n)$ is given by
\be\ba
\label{centr-stop-time}&p(n+1)=p(u^n=c)\triangleq p(u_1=c,...,u_n=c),\\
&\sum_{n=1}^\infty p(u^n=c)=1,\\
&p(u_n=j|u_t=j'\neq c)=\delta_{jj'},~~~~\txt{for all}~~t<n.
\ea\ee
Note that $p(n+1)$ is not the same as $p(x_{n+1})$. The last set of constraints in (\ref{centr-stop-time}) has the explicit form
\be\ba
&p(u_n=c|u_t=0)=p(u_n=c|u_t=1)=0,~~~~\txt{for all}~~t<n,\\
&p(u_n=0|u_t=0)=p(u_n=1|u_t=1)=1,~~~~\txt{for all}~~t<n,\\
&p(u_n=0|u_t=1)=p(u_n=1|u_t=0)=0,~~~~\txt{for all}~~t<n.\nn
\ea\ee
These natural \emph{quickest detection constraints} greatly reduce the number of nontrivial optimization variables in the objective function we will consider. In addition to (\ref{centr-stop-time}), and without loss of generality, we can further simplify computation by setting
\be
\label{centr-grounding}p(u=0|x_1)=p(u=1|x_1)=0,
\ee
i.e., we may assume that at least $2$ samples must be taken. It seems worthwhile to mention that in this section we shall be involved with the minimization of a \emph{convex} function of the form
\bea
f(x,y,z)=\sum_{n=1}^\infty\big(a_nx_n+b_ny_n+c_nz_n\big)y_1y_2\cdots y_{n-1},\nn
\eea
where $a_n,b_n,c_n$ are constant positive real numbers, and $0\leq x_n,y_n,z_n\leq 1$ for all $n=1,2,\cdots$, i.e., $x,y,z\in [0,1]^{\mathbb{Z}_+}$.

Given a \emph{false alarm} level $P_f=p(u_n=1|H_0)\leq\al$ and a \emph{missed detection} level $P_m=p(u_n=0|H_1)\leq\beta$, our objective is to minimize the expected number of samples
\be\ba
\label{centr-sequential-objective}&S=\sum_{n=1}^\infty(n+1)p(n+1)=\sum_{n=1}^\infty (n+1)p(u^n=c)\\
&~~~~=\sum_{n,x^n,i} (n+1)p(u^n=c|x^n)p_i(x^n)\pi_i\\
&~~~~=\sum_{n,x^n,i} (n+1)\prod_{t=1}^np(u_t=c|x^t,u^{t-1}=c)~p_i(x^n)\pi_i\\
&~~~~=\sum_{n,x^n,i} (n+1)\prod_{t=1}^nq(u_t=c|x^t)~p_i(x^n)\pi_i\\
&~~~~\sr{(\ref{centr-stop-time})}{=}\sum_{n=1}^\infty (n+1)p(u_n=c),
\ea\ee
where $\pi_i=p(H_i)$, and
\be\ba
\label{sequential-distribution}&q(u_t=j|x^t)\triangleq p(u_t=j|x^t,u^{t-1}=c)\\
&~~~~=p(u_t=j|x^t,u_1=c,...,u_{t-1}=c).
\ea\ee
Thus we have the optimization problem
\be\ba
\label{centr-problem}&\txt{minimize}~~S\\
&\txt{subject to}~~\{p_0(u_n=1)\leq\al,~p_1(u_n=0)\leq\beta\}_n,\\
&\hspace{2cm}n=1,2,\cdots,~~\sum_{n=1}^\infty p(u^n=c)=1,
\ea\ee
where, by (\ref{centr-stop-time}), we have (\emph{up to unimportant terms})
\be\ba
\label{centr-terms-expansion}&p_i(u_n=j)=\sum_{x^n,u^{n-1}}p_i(u_n=j,x^n,u^{n-1})\\
&~~=\sum_{x^n}p_i(x^n)q(u_n=j|x^n)\prod_{t=1}^{n-1}q(u_t=c|x^t),\\
&~~=p_i(u_n=j,u^{n-1}=c),
\ea\ee
and the optimization variables are
\be
\label{centr-opt-vars}\left\{q(u_n=j|x^n)\right\}_{n,j,x^n}.\nn
\ee
The Lagrangian for the problem (\ref{centr-problem}) is
\be\ba
&L=\sum_n\bigg((n+1)p(u^n=c)+\ld_{n0}\big[\al-p_0(u_n=1)\big]\nn\\
&~~~~+\ld_{n1}\big[\beta-p_1(u_n=0)\big]+\ld\big[s_n-p(u^n=c)\big]\bigg),\nn
\ea\ee
where $\ld_{n0},\ld_{n1}\leq 0,\ld\in\Real$, $p(u^n=c)=p_0(u^n=c)\pi_0+p_1(u^n=c)\pi_1$, and $s_n$ is any sequence of numbers such that $\sum s_n=1$. We can perform the rescaling $\ld_{ni}\ra (n+1)\ld_{ni},\ld\ra (n+1)\ld$ without loss of generality. Thus we can write
\be\ba
\label{centr-lagrangian}&L=\sum_n(n+1)\bigg(p(u^n=c)+\ld_{n0}\big[\al-p_0(u_n=1)\big]\\
&~~~~+\ld_{n1}\big[\beta-p_1(u_n=0)\big]+\ld\big[s_n-p(u^n=c)\big]\bigg).
\ea\ee

\begin{thm}\label{centr-thm}
Suppose the observations are conditionally independent so that $p_i(x^n)=\prod_{t=1}^np_i(x_t)$ for all $n=1,2,\cdots$. Then the optimal decision rule for the test with the Lagrangian (\ref{centr-lagrangian}) as objective is
\bea
\label{centr-dec-rule}q_{\txt{opt}}(u_n=j|x^n)=I_{R_{u_n=j}}(x^n),~~j\in\{0,c,1\},
\eea
where the decision regions are given by
\be\ba
\label{centr-dec-regions}& R_{u_n=0}=\left\{x^n:{p_1(x^n)\over p_0(x^n)}<A_n\right\},\\
& R_{u_n=c}=\left\{x^n:A_n<{p_1(x^n)\over p_0(x^n)}<B_n\right\},\\
& R_{u_n=1}=\left\{x^n:B_n<{p_1(x^n)\over p_0(x^n)}\right\}.
\ea\ee
The threshold parameters $A_n,B_n$, which are independent of the observations $x^n$, are given by (\ref{c-thresholds}).
\end{thm}
\begin{IEEEproof}
See Appendix \ref{centr-thm-proof}
\end{IEEEproof}
\begin{rmks}~

~~0) Without the conditional independence assumption, the two-threshold likelihood ratio test obtained in the theorem will be suboptimal in general.

~~1)  The result of the theorem shows that Wald's \emph{sequential probability ratio test} (SPRT) in \cite{wald-45} indeed minimizes the expected stopping time provided the thresholds are made to depend on $n$ accordingly.

~~2) Note that in the theorem, it is assumed that the sum $S$ in (\ref{centr-sequential-objective}) converges, a requirement we will establish in Theorem \ref{cst-convergence}. It was also shown in \cite{wald-44} that this is the case if and only if the observations have nonzero variance.

~~3) Notice that the function $S$ in (\ref{centr-sequential-objective}) is a special case of an objective function of the form

{\small
\be\ba
&S=E[C_{n,u^n,h}]=\sum_{n,u^n,h}C_{n,u^n,h}p(n,u^n,h)\nn\\
&~=\sum_{n,h}C_{n,(c,...,c),h}p(n,u^n=c,h)+\sum_{n,u^n\neq c,h}C_{n,u^n,h}p(n,u^n,h)\nn\\
&~=\sum_{n,h}(n+1)p(u^n=c,h)+\sum_{n,u^n\neq c,h}C_{n,u^n,h}p(n,u^n,h)\nn
\ea\ee
}where the 1st term is the expected number of observations, and the 2nd term is the usual expected cost of deciding one of the two hypotheses at various time instances.

~~4) If $\tau$ is the \emph{stopping time} as a random variable, then the \emph{stopping rule} is given by
\be\ba
&p(\tau=n|x^n) \triangleq q_{\txt{opt}}(u_n\neq c|x^n)=1-q_{\txt{opt}}(u_n=c|x^n)\\
&~~~~=1-I_{R_{u_n=c}}(x^n).
\ea\ee

~~5) In order to determine the optimal value $S_{\txt{opt}}$ of the objective function, we can either directly minimize the Lagrangian function $L(\ld,\ld_{0n},\ld_{1n},A_n,B_n)$, or simply minimize the sample number function $S(A_n,B_n)$ subject to the constraints $p(u_n=1|H_0)\leq\al$, $p(u_n=0|H_1)\leq\beta$, which by (\ref{centr-terms-expansion}) can now be written as
\be\ba
\label{centr-error-constraints-opt}&p_0\left(R_{u_n=1}~\cap~\mathop{\cap}\limits_{t=1}^{n-1}\widehat{R}_{u_t=c}\right)\leq\al,\\
&p_1\left(R_{u_n=0}~\cap~\mathop{\cap}\limits_{t=1}^{n-1}\widehat{R}_{u_t=c}\right)\leq\beta,
\ea\ee
where $\widehat{R}_{u_t=c}=\Real^{n-t}\times R_{u_t=c}$. This optimal value of $S$ in terms of probabilities of the decision regions is
\be\ba
\label{optimal-sample-number}&S_{\txt{opt}}=\sum_{n,x^n,i} (n+1)\prod_{t=1}^nq_{\txt{opt}}(u_t=c|x^t)~p_i(x^n)\pi_i\\
&~~~~=\sum_{n,x^n,i}(n+1)\prod_{t=1}^nI_{R_{u_t=c}}(x^t)~p_i(x^n)\pi_i\\
&~~~~=\sum_{n,x^n,i}(n+1)I_{\cap_{t=1}^n\widehat{R}_{u_t=c}}(x^n)~p_i(x^n)\pi_i\\
&~~~~=\sum_{n,i}(n+1)~p_i\left(\mathop{\cap}\limits_{t=1}^n\widehat{R}_{u_t=c}\right)\pi_i.
\ea\ee

6) Note that for $S$ to be a true expectation in the sense of probability theory, the summand in $S$ should include a \emph{global} normalization constant $K$, which for convenience, we have simply set equal to $1$ by normalizing the defining constraint $\sum p(u^n=c)=1$ in (\ref{centr-stop-time}). That is, we should actually write $S=\sum(n+1){1\over K}p(u^n=c)$ with the defining constraint in the form $\sum p(u^n=c)=K$, where $K$ is by definition independent of the optimization variables (and hence independent of the thresholds). The actual value of $K$, however, will depend on the problem at hand. A natural choice for $K$ is
\be
K={1\over \sum p_{\txt{opt}}(u^n=1)},\nn
\ee
which is to be computed only after the optimal point must have been found. This implies the objective function $S=\sum(n+1){1\over K}p(u^n=c)$ is truly an expected sample number only at the optimal point. This is sufficient since we are only interested in the optimal sample number.

~~7) If the constraint $\sum p(u^n=c)=1$ is not included in the optimization problem (\ref{centr-problem}), the resulting threshold structure is the same. Therefore, this is not an essential constraint for the purpose of optimization. However, the fact that its inclusion does not affect the threshold structure serves as a consistency check, showing that $S$ is a well defined expectation in that it is consistent with the basic rules of probability.

~~8) The distribution of the optimal stopping time can be studied using its characteristic function
\be
\psi_{S_{\txt{opt}}}(z)=E[e^{nz}]=\sum_{n,i} e^{nz}~p_i\left(\mathop{\cap}\limits_{t=1}^n\widehat{R}_{u_t=c}\right)\pi_i.\nn
\ee
\end{rmks}

\begin{thm}\label{cst-convergence}
If the observations are conditionally independent and the likelihood ratio has nontrivial dependence on data, then the quantity $S_{\txt{opt}}$ in (\ref{optimal-sample-number}) converges.
\end{thm}
\begin{IEEEproof}
Consider a \emph{sample-by-sample (sbs) test} whose decision rule (not necessarily optimal)
\be\ba
& q^{\txt{(sbs)}}(u_n=j|x_n)=I_{R^{\txt{(sbs)}}_{u_n=j}}(x_n),~~~~u_n=\gamma_n(x_n,u^{n-1}),\\
\label{centr-SBS-region}&R^{\txt{(sbs)}}_{u_n=0}=\left\{x_n:{p_1(x_n)\over p_0(x_n)}<A\right\},\\
&R^{\txt{(sbs)}}_{u_n=c}=\left\{x_n:A\leq {A_n\over B_{n-1}}<{p_1(x_n)\over p_0(x_n)}<{B_n\over A_{n-1}}\leq B\right\},\\
&R^{\txt{(sbs)}}_{u_n=1}=\left\{x_n:B<{p_1(x_n)\over p_0(x_n)}\right\}
\ea\ee
is determined in the same way as the full sample rule given by (\ref{centr-dec-rule}) and (\ref{centr-dec-regions}). Here, a \textbf{sample-by-sample (sbs) test} is one where the decision $u_t$ depends on previous observations $x^{t-1}=(x_1,...,x_{t-1})$ only through the previous decisions $u^{t-1}=(u_1,...,u_{t-1})$, so that $u_t=u_t(x_t,u^{t-1})$. Note that the defining conditions (\ref{centr-stop-time}) for stopping time are still required for the sample-by-sample test, and so details of the decision procedure are the same as for the full-sample test.

Let $S_{\txt{sbs}}(A_n,B_n)$ denote the objective function corresponding to the sample-by-sample test, i.e.,
\be\ba
\label{centr-SBS-objective}&S_{\txt{sbs}}(A_n,B_n)=\sum_{n,x^n,i} (n+1)\prod_{t=1}^nq^{\txt{(sbs)}}(u_t=c|x_t)~p_i(x^n)\pi_i\\
&~~~~=\sum_{n,x^n,i}(n+1)\prod_{t=1}^nI_{R^{\txt{(sbs)}}_{u_t=c}}(x_t)~p_i(x^n)\pi_i\\
&~~~~=\sum_{n,x^n,i}(n+1)~I_{\prod_{t=1}^nR^{\txt{(sbs)}}_{u_t=c}}(x^n)~p_i(x^n)\pi_i\\
&~~~~=\sum_{n,i}(n+1)~p_i\left(\prod_{t=1}^nR^{\txt{(sbs)}}_{u_t=c}\right)\pi_i.
\ea\ee
Then we know $S_{\txt{opt}}(A_n,B_n)\leq S_{\txt{sbs}}(A_n,B_n)$. Moreover,
\be
\label{centr-SBS-bound}S_{\txt{opt}}(A_n,B_n)\leq S_{\txt{sbs}}(A,B),
\ee
for any thersholds $(A,B)$. For simplicity, we will consider only thresholds $(A,B)$ that do not depend on $n$.

Since we assume the observations are conditionally independent, i.e., $p_i(x^n)=\prod_{k=1}^np_i(x_k)$, by (\ref{optimal-sample-number}), (\ref{centr-SBS-objective}), and (\ref{centr-SBS-bound}) we obtain
\be\ba
&S_{\txt{opt}}=\sum_{n,x^n,i} (n+1)\prod_{t=1}^nI_{R_{u_t=c}}(x^t)~p_i(x^n)\pi_i\\
&~~~~\leq S_{\txt{sbs}}=\sum_{n,x^n,i} (n+1)\prod_{t=1}^n\left(I_{R^{\txt{(sbs)}}_{u_t=c}}(x_t)p_i(x_t)\right)\pi_i\\
\label{centr-SBS-CIO-bound}&~~~~\leq U=\sum_{n,i} (n+1)e^{-nL_i}\pi_i=\sum_{n,i}\left(-{d\over dL_i}+1\right)e^{-nL_i}\pi_i\\
&~~~~=\sum_i{2e^{L_i}-1\over (e^{L_i}-1)^2}\pi_i,
\ea\ee
where $L_i=-\log~p_i\big(A<p_1(x)/p_0(x)<B\big)$.
\end{IEEEproof}

\begin{rmks}~\\
1)~~Suppose we drop the conditional independence assumption in the theorem. Then the suboptimal sample number $S_{\txt{sopt}}$ due to the underlying suboptimal three-threshold test converges under the following condition. Fix $\vep>0$ and suppose the limit
\bea
C_i=\lim_{n\ra\infty}~C_i(x^n),~~C_i(x^n)={1\over n}\log{p_i(x^n)\over \prod_{k=1}^np_i(x_k)},\nn
\eea
exists. Then there is an integer $N$ such that $|C_i(x^n)-C_i|\leq\vep$ for all $n\geq N$. Thus if $n\geq N$, we have
\bea
p_i(x^n)=e^{nC_i(x^n)}\prod_{k=1}^np_i(x_k)\leq e^{n(C_i+\vep)}\prod_{k=1}^np_i(x_k).\nn
\eea
If in addition the correlation rate $C_i$ is such that $L_i>C_i+\vep$, then $S_{\txt{opt}}$ converges, with an upper bound
\be\ba
&S_{\txt{sopt}}~\leq~S_N+\sum_{n>N,i} (n+1)e^{-n(L_i-C_i-\vep)}\pi_i\nn\\
&~~~~=S_N+\sum_i\pi_i\left(-{d\over dt}+1\right){e^{-(N+1)t}\over 1-e^{-t}}\bigg|_{t=L_i-C_i-\vep}\nn\\
&~~~~=S_N+\sum_i\pi_i{N(e^t-1)+2e^t-1\over e^{Nt}(e^t-1)^2}\bigg|_{t=L_i-C_i-\vep},\nn
\ea\ee
where
\be
S_N=\sum_{n=1}^N\sum_{x^n,i} (n+1)\prod_{t=1}^nI_{R_{u_t=c}}(x^t)~p_i(x^n)\pi_i,\nn
\ee
and the quantity on the right hand side of the inequality is viewed as the objective function for a suboptimal test in which sample-by-sample processing is performed for $n\geq N$.

2)~~Since the thresholds $A,B$ in the upper bound (\ref{centr-SBS-CIO-bound}) are independent of $n$, the error constraints $p_0(u_n=1)\leq\al$, $p_1(u_n=0)\leq\beta$ for the sample-by-sample test take the form
\be
\label{centr-SBS-constraints}p_0\big(B<p_1(x_t)/p_0(x_t)\big)\leq\al,~~p_1\big(p_1(x_t)/p_0(x_t)<A\big)\leq\beta.
\ee

3) Henceforth in this section, we will no longer indicate the dependence of the thresholds on $n$, since we will mostly be dealing with the upper bound in (\ref{centr-SBS-CIO-bound}).
\end{rmks}

\begin{example}[Constant signal in Gaussian noise]\label{S-test-example}
Assuming the observations are identical and conditionally independent, the hypotheses are
\be
H_i:~x_n\sim \txt{Normal}(i,\sigma^2),~~i=0,1,~~n=1,2,\cdots.\nn
\ee
The relevant probabilities in terms of $Q(x)={1\over\sqrt{2\pi}}\int_x^\infty e^{-{t^2\over 2}}dt$ are
\be\ba
& p_i\big(B<p_1(x)/p_0(x)\big)=Q\big((b-i)/\sigma\big),\nn\\
& p_i\big(A<p_1(x)/p_0(x)<B\big)\nn\\
&~~~~~~=Q\big((a-i)/\sigma\big)-Q\big((b-i)/\sigma\big),\nn\\
& p_i\big(p_1(x)/p_0(x)<A\big)=1-Q\big((a-i)/\sigma\big),\nn
\ea\ee
where
\be
a=\sigma^2\log A+1/2,~~~~b=\sigma^2\log B+1/2.\nn
\ee
The constraints (\ref{centr-SBS-constraints}) become
\be
\label{centr-SBS-constraints2}Q\big(b/\sigma\big)\leq\al,~~~~1-Q\big((a-1)/\sigma\big)\leq\beta,
\ee
while the quantity $L_i$ in (\ref{centr-SBS-CIO-bound}) becomes
\be
L_i=-\log\big[Q\big((a-i)/\sigma\big)-Q\big((b-i)/\sigma\big)\big].\nn
\ee
The bound in (\ref{centr-SBS-CIO-bound}) gives
\be\ba
S_{\txt{opt}}\leq \sum_if(\Delta_i)\pi_i,\nn
\ea\ee
where~ $\Delta_i=Q\big((a-i)/\sigma\big)-Q\big((b-i)/\sigma\big)$ and
\be
\label{base-function}f(x)=x(2-x)/(1-x)^2={1\over(1-x)^2}-1.
\ee
Numerical results for this example are shown in Fig \ref{Centralized-SN}. The shapes of the graphs, and the numbers on them as well, are observed to be independent of the choice of prior probability $\pi$.

\begin{figure}[H]
\centering
\scalebox{0.6}{\includegraphics[width=15cm,height=9cm]{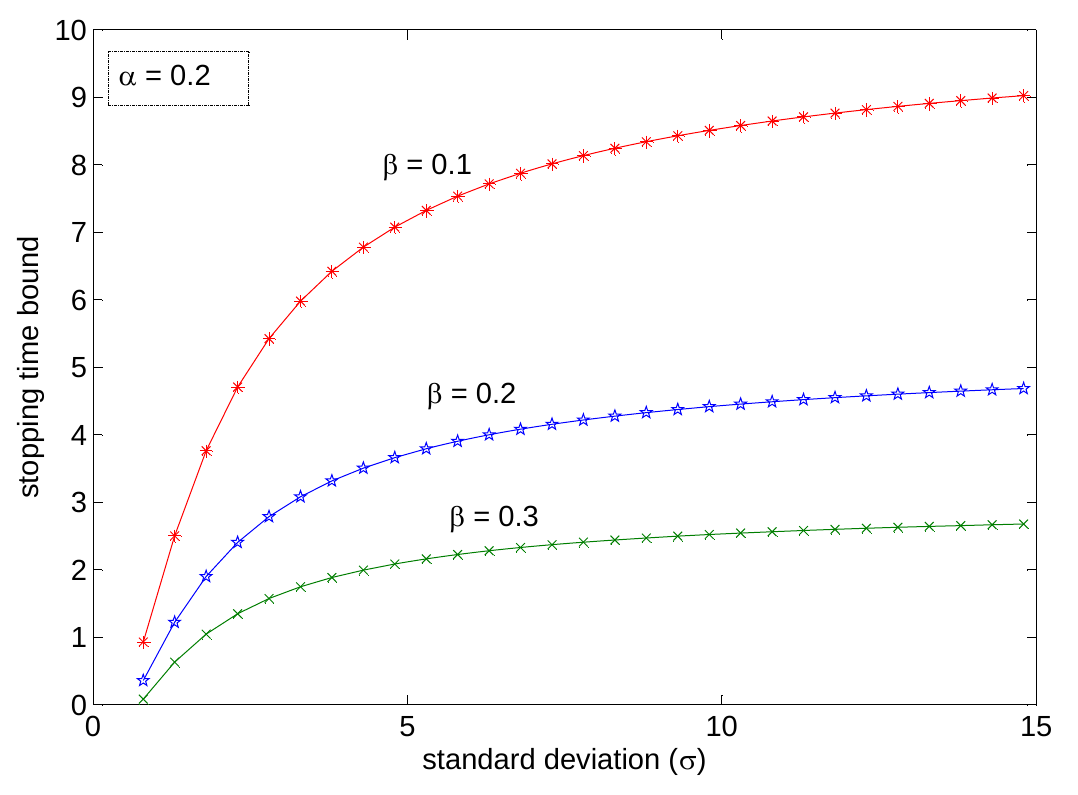}}\\
\caption{One-sensor centralized test: Dependence of the stopping time bound on standard deviation and missed detection level $\beta$.}\label{Centralized-SN}
\end{figure}
\end{example}

\subsection{Savings in number of observations}
The number of observations saved by performing the sequential test is $\Delta n=N(\al,\beta)-S_{\txt{opt}}$, where $n=N(\al,\beta)$ is a solution of the system (with $\widehat{R}_{u_t=c}$ as in (\ref{centr-error-constraints-opt}))
\be\ba
&p_0\left(R_{u_n=1}~\cap~\mathop{\cap}\limits_{t=1}^{n-1}\widehat{R}_{u_t=c}\right)\big|_{A=B}=\al,\\
&p_1\left(R_{u_n=0}~\cap~\mathop{\cap}\limits_{t=1}^{n-1}\widehat{R}_{u_t=c}\right)\big|_{A=B}=\beta.\nn
\ea\ee
Because the sample mean $\bar{x}_n={1\over n}\sum_{i=1}^nx_i$ determines the likelihood ratio $p_1(x^n)/p_0(x^n)$ and is distributed according to $\bar{x}\sim \txt{Normal}(\mu,\sigma^2/n)$, solving the above system for $A=B$ and $n$ is \emph{approximately} equivalent to solving the system
\bea
\label{usual-constr}Q(\sqrt{n}~\!t/\sigma)=\al,~~~~1-Q\big(\sqrt{n}~\!(t-1)/\sigma\big)=\beta
\eea
for $t$ and $n=N(\al,\beta)$. The sequential test thresholds $(a_0,b_0)$ corresponding to $n_0=N(\al,\beta)$ can then be obtained by solving for $a$ and $b$ in the system
\bea
Q(\sqrt{n_0}~\!b/\sigma)=\al,~~~~1-Q\big(\sqrt{n_0}~\!(a-1)/\sigma\big)=\beta.\nn
\eea
Thus the savings in the number samples are
\be
\Delta n=n_0-S_{\txt{opt}}(a_0,b_0).\nn
\ee

Note that in general, the constraints (\ref{usual-constr}) are not compatible with the sample-by-sample test constraints (\ref{centr-SBS-constraints2}) used to investigate the behavior of the upper bound $S_{\txt{sbs}}$ of $S_{\txt{opt}}$. Nevertheless, at $\sigma=1$ we have $S_{\txt{sbs}}\approx 1$ as shown in Fig \ref{Centralized-SN}, in which case, the identification $n\approx S_{\txt{sbs}}\approx 1$ implies that (\ref{centr-SBS-constraints2}) and (\ref{usual-constr}) are compatible, provided we view the sequential thresholds $a,b$ in (\ref{centr-SBS-constraints2}) as approximations of the static threshold $t$ in (\ref{usual-constr}). Thus, we may estimate the corresponding number of samples for the static (i.e., non-sequential) test as follows. Let the thresholds $(a_0,b_0)$ be obtained by solving for $(a,b)$ in (\ref{centr-SBS-constraints2}) with $\sigma=1$, where $a_0$ and $b_0$ are viewed as approximations of the threshold $t$ in (\ref{usual-constr}), again with $\sigma=1$. Next, let $t={a_0+b_0\over 2}$. Consider the modification
\be
\label{centr-SBS-constraints22} 1-Q(\sqrt{n_s}~\!(t-1))=\beta
\ee
of the second equation in (\ref{centr-SBS-constraints2}) with $\sigma=1$, where $n_s$ is the number of samples required by the static likelihood ratio test to reach the accuracy level $\beta$, and with $a$ replaced by $t$. Then $\delta n=n_s-U\approx n_s-1$ ($U$ denoting the upper bound on $S_{\txt{sbs}}$) is the excess number of samples required by the static test at $\sigma=1$. Numerical computation (not presented here) shows that $\delta n\approx 1$ (i.e., ${\delta n\over n_s}\approx {1\over 2}$). This is compatible with the observation made in \cite{wald-45} that about $50\%$ sampling time can be saved by using the sequential likelihood ratio test (SLRT).

Before discussing the distributed sequential test, we first present numerical results for the centralized two-sensor SLRT.

\subsection{Centralized version of the two-sensor distributed SLRT}\label{cvdt-section}
\begin{figure}[H]
\centering
\scalebox{0.6}{\includegraphics[width=15cm,height=9cm]{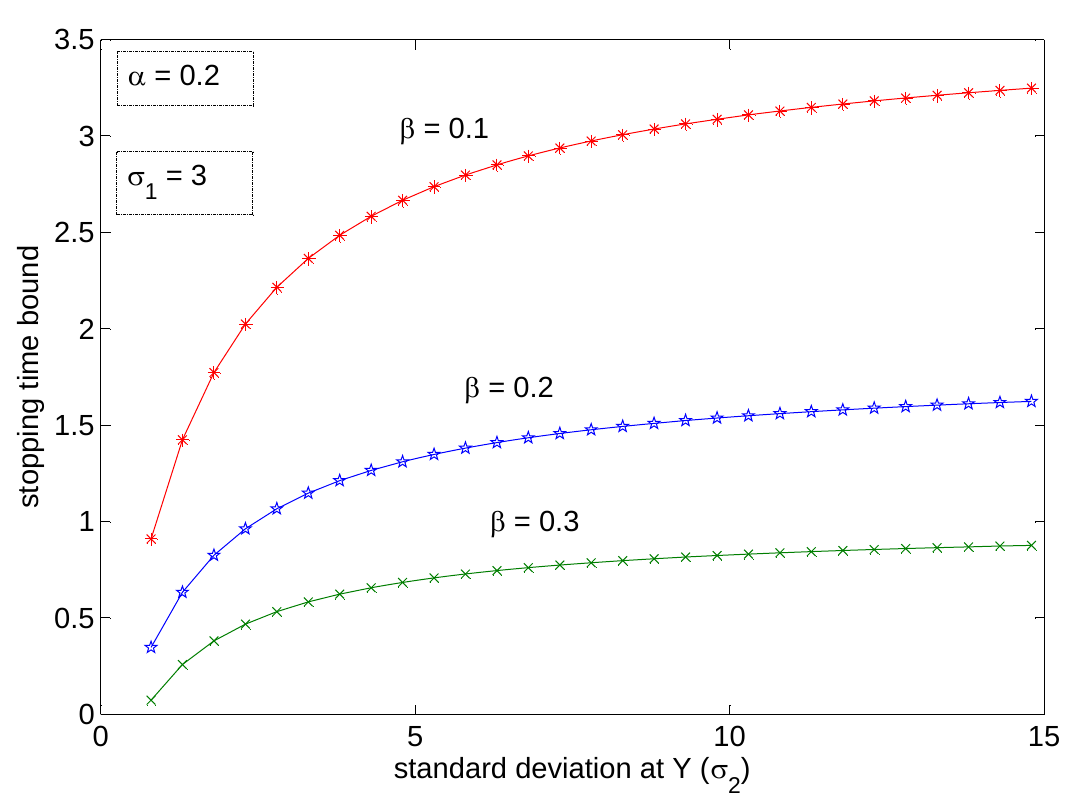}}\\
\caption{Two-sensor centralized test: Dependence of the stopping time bound on standard deviation at Y and missed detection level $\beta$.}\label{2Sensor-Centralized-SN}
\end{figure}

We again consider Example \ref{S-test-example}. Let $\{X,Y\}$ be a network of two sensors whose observations are available at the same location. The sequential test for this network is a centralized version of the sequential test for the distributed network in Fig \ref{2-tandem}. The relevant probabilities for this centralized test in terms of $Q$-functions are as follows. With $l(x,y)=p_1(x,y)/p_0(x,y)$,
\be\ba
& p_i\big(B<l(x,y)\big)\nn\\
&~~={1\over\sqrt{2\pi\sigma_2^2}}\int_{-\infty}^\infty Q\left({b-(\sigma_1/\sigma_2)^2y-i\over\sigma_1}\right)e^{-{(y-i)^2\over2\sigma_2^2}}dy,\nn\\
& p_i\big(l(x,y)<A\big)\nn\\
&~~=1-{1\over\sqrt{2\pi\sigma_2^2}}\int_{-\infty}^\infty Q\left({a-(\sigma_1/\sigma_2)^2y-i\over\sigma_1}\right)e^{-{(y-i)^2\over2\sigma_2^2}}dy,\nn\\
& p_i\big(A<l(x,y)<B\big)=1-p_i\big(B<l(x,y)\big)\nn\\
&~~~~~~-p_i\big(l(x,y)<A\big),\nn
\ea\ee
where
\be\ba
a=\sigma_1^2\log A+{\sigma_1^2+\sigma_2^2\over 2\sigma_2^2},~~b=\sigma_1^2\log B+{\sigma_1^2+\sigma_2^2\over 2\sigma_2^2}.\nn
\ea\ee
The constraints (\ref{centr-SBS-constraints}) become
\be\ba
&{1\over\sqrt{2\pi\sigma_2^2}}\int_{-\infty}^\infty Q\left({b-(\sigma_1/\sigma_2)^2y\over\sigma_1}\right)e^{-{y^2\over2\sigma_2^2}}dy\leq\al,\nn\\
&1-{1\over\sqrt{2\pi\sigma_2^2}}\int_{-\infty}^\infty Q\left({a-(\sigma_1/\sigma_2)^2y-1\over\sigma_1}\right)e^{-{(y-1)^2\over2\sigma_2^2}}dy\leq\beta.\nn
\ea\ee
The bound in (\ref{centr-SBS-CIO-bound}) gives $S_{\txt{opt}}\leq\sum_if(\Delta_i)\pi_i$, where $f$ is the function in (\ref{base-function}), and
\be\ba
\Delta_i=1-p_i\big(B<l(x,y)\big)-p_i\big(l(x,y)<A\big).\nn
\ea\ee
The results of numerical computation are shown in Fig \ref{2Sensor-Centralized-SN}.

\section{The Distributed Sequential Test: Two sensors}\label{dst-section}
For simplicity, we shall work with a two-sensor tandem network. The test for this simple network contains the basic essential features of the test for a general network in the form of an acyclic directed graph to be considered in Section \ref{generalize}. Thus, the test described here represents the distributed test in its simplest nontrivial form.

\begin{figure}[H]
\centering
\scalebox{0.4}{\includegraphics[width=18cm,height=10cm]{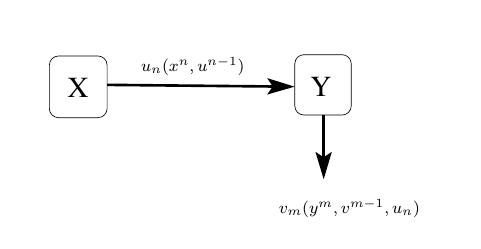}}\\
\caption{Two-sensor tandem network}\label{2-tandem}
\end{figure}

In the network of two sensors $\{X,Y\}$ in Fig \ref{2-tandem}, the observations of $X$ are $x^n=(x_1,...,x_n)$ and those of $Y$ are $y^m=(y_1,...,y_m)$. We assume $X$ forwards its decision $u_n=\gamma_n(x^n,u^{n-1})$ to $Y$, and $Y$ makes the final decision
\be
v_m=\left\{
            \begin{array}{ll}
              \rho_m(y^m,v^{m-1}), & m<n \\
              \rho_m(y^m,v^{m-1},u_n), & n\leq m
            \end{array}
          \right\},\nn
\ee
i.e., $Y$ is the fusion center. For simplicity, we will assume that $Y$ begins sampling only after receiving the decision of $X$, so that $v_m=\rho_m(y^m,v^{m-1},u_n)$, where we impose the timing constraint $n\leq m$. For notational convenience however, $\sum_{m,n\leq m}$ will be written simply as $\sum_{m,n}$ with the timing constraint understood.

As for the centralized test, we consider \emph{two} hypotheses
\bea
H_i:(x^n,y^m)\sim p_i(x^n,y^m)=p(x^n,y^m|H_i),~~i=0,1,
\eea
and \emph{ternary} decision rules
\be\ba
&\gamma_n:(x^n,u^{n-1})\in\X^n\times\U^{n-1}\\
&~~~~\longmapsto u_n=\gamma_n(x^n,u^{n-1})\in \U_n=\{0,c,1\},\nn\\
&\rho_m:(y^m,v^{m-1},u_n)\in\Y^m\times\V^{m-1}\times\U_n\nn\\
&~~~~~~~~\longmapsto v_m=\rho_m(y^m,v^{m-1},u_n)\in \V_m=\{0,c,1\}.\nn
\ea\ee
Here, $u_n=0$ or $v_m=0$ is acceptance of $H_0$, $u_n=1$ or $v_m=1$ is acceptance of $H_1$, and $u_n=c$ or $v_m=c$ is rejection of both $H_0$ and $H_1$ in favor of continued sampling. At $X$, the next sample $x_{n+1}$ is taken if and only if
\bit
\item $\gamma_t(x^t,u^{t-1})=c$~ for all $1\leq t\leq n$.
\eit
At $Y$, for any given $n$ (representing the stopping time of X), the next sample $y_{m+1}$ is taken if and only if the following hold:
\bit
\item $\rho_s(y^s,v^{s-1},u_n)=c$~ for all $n\leq s\leq m$,
\eit
where an additional constraint, $\rho_s(y^s,v^{s-1})=c$~ for all $1\leq s<n$, would be needed if we drop the assumption that Y begins sampling only after receiving the decision of X.

Therefore, the sample size $n=g_n(u^{n-1})$ is a random variable whose distribution $p(n)$ satisfies (\ref{centr-stop-time}). Similarly, for each $n=1,2,\cdots$, the sample size $m=h_m(v^{m-1},u_n)$ is a random variable whose distribution $p(m)$ satisfies
\be\ba
\label{distr-stop-time}&p(m+1)=p(v^m=c)\triangleq p(v_1=c,...,v_m=c),\\
&\sum_{m=1}^\infty p(v^m=c)=1,\\
&p(v_m=j|v_s=j'\neq c)=\delta_{jj'},~~~~\txt{for all}~~n\leq s<m.
\ea\ee

At the fusion center $Y$, given a \emph{false alarm} level $p_0(v_m=1)=p(v_m=1|H_0)\leq\al$ and a \emph{missed detection} level $p_1(v_m=0)=p(v_m=0|H_1)\leq\beta$, our objective is to minimize the expected number of samples
\be\ba
\label{distr-sequential-objective}&S=\sum_{m=1}^\infty(m+1)p(m+1)=\sum_{m=1}^\infty (m+1)p(v^m=c)\\
&~~~~=\sum_{m,y^m,n,u^n,x^n,i} (m+1)p(v^m=c,y^m,n,u^n,x^n,H_i)\\
&~~~~=\sum (m+1)p(v^m=c|y^m,u_n)p(n|x^n)p(u^n|x^n)\\
&~~~~~~~~\times p_i(x^n,y^m)\pi_i\\
&~~~~=\sum (m+1)\prod_{s=1}^mp(v_s=c|y^s,v^{s-1}=c,u_n)\\
&~~~~~~~~\times p(n|x^n)\prod_{t=1}^np(u_t|x^t,u^{t-1})~p_i(x^n,y^m)\pi_i\\
&~~~~=\sum (m+1)\prod_{s=1}^mq(v_s=c|y^s,u_n)\\
&~~~~~~~~\times p(n|x^n)\prod_{t=1}^np(u_t|x^t,u^{t-1})~p_i(x^n,y^m)\pi_i\\
&~~~~\sr{(\ref{distr-stop-time})}{=}\sum_{m=1}^\infty (m+1)p(v_m=c),
\ea\ee
where $q(v_s|y^s,u_n)=p(v_s|y^s,v^{s-1}=c,u_n)$. By definition, ~$p(n|x^n)=p(u^{n-1}=c|x^n)=1$~ because at time $n-1$, the next sample $x_n$ is taken if and only if $u^{n-1}=c$. Therefore,
\be\ba
\label{tandem-SO}&S=\sum_{m,y^m,n\leq m,u^n,x^n,i} (m+1)\prod_{s=1}^mq(v_s=c|y^s,u_n)\\
&~~~~~~~~\times\prod_{t=1}^np(u_t|x^t,u^{t-1})~p_i(x^n,y^m)\pi_i.
\ea\ee

Thus, we have the optimization problem
\be\ba
&\txt{minimize}~S\\
&\txt{subject to}~\{p_0(v_m=1)\leq\al,~p_1(v_m=0)\leq\beta\}_m,\\
&~~~~m=1,2,...,~~~~\sum_{m=1}^\infty p(v^m=c)=1,
\ea\ee
where the optimization variables are
\bea
\{q(u_n=j|x^n),q(v_m=j'|y^m,u_n),\}_{n,j,x^n,m,j',y^m,u_n}.\nn
\eea
and we again have the \emph{essential-terms expansion} (\ref{centr-terms-expansion}) for $p_i(u_n=j)$ due to (\ref{centr-stop-time}), along with a similar expression for $p_i(v_m=j)$ due to (\ref{centr-stop-time}) and (\ref{distr-stop-time}), which is (up to unimportant terms) given by

{\small
\be\ba
\label{distr-terms-expansion}&p_i(v_m=j)=\sum_{v^{m-1},y^m,n,u^n,x^n}p_i(v_m=j,v^{m-1},y^m,n,u^n,x^n)\\
&~~=\sum_{y^m,v^{m-1},n,x^n,u^n}p(v_m=j,v^{m-1}|y^m,u_n)\\
&~~~~\times p(u^n|x^n)p_i(x^n,y^m)\\
&~~=\sum_{n,x^n,y^m,u_n}q(v_m=j|y^m,u_n)\prod_{s=1}^{m-1}q(v_s=c|y^s,u_n)\\
&~~~~\times q(u_n|x^n)\prod_{t=1}^{n-1}q(u_t=c|x^t)~p_i(x^n,y^m)\\
&~~=\sum_{n,u_n}p_i(v_m=j,v^{m-1}=c,u_n,u^{n-1}=c)\\
&~~=p_i(v_m=j,v^{m-1}=c).
\ea\ee
}

The Lagrangian for the above problem is
\be\ba
&L=\sum_m\bigg((m+1)p(v^m=c)+\ld_{m0}\big[\al-p_0(v_m=1)\big]\\
&~~~~+\ld_{m1}\big[\beta-p_1(v_m=0)\big]+\ld\big[s_m-p(v^m=c)\big]\bigg),\nn
\ea\ee
where~ $\ld_{m0},\ld_{m1}\leq 0,\ld\in\Real$,~ $p(v^m=c)=p_0(v^m=c)\pi_0+p_1(v^m=c)\pi_1$,~ and $s_m$ is any sequence of numbers such that $\sum s_m=1$. We can perform the rescaling $\ld_{mi}\ra(m+1)\ld_{mi}$, $\ld\ra (m+1)\ld$ without loss of generality. Thus we can write
\be\ba
\label{distr-lagrangian}&L=\sum_m(m+1)\bigg(p(v^m=c)+\ld_{m0}\big[\al-p_0(v_m=1)\big]\\
&~~~~+\ld_{m1}\big[\beta-p_1(v_m=0)\big]+\ld\big[s_m-p(v^m=c)\big]\bigg),\\
&p(v^m=c)=p_0(v^m=c)\pi_0+p_1(v^m=c)\pi_1.
\ea\ee

\begin{thm}\label{distr-thm}
Suppose the observations for the network in Fig \ref{2-tandem} are conditionally independent so that
\be
p_i(x^n,y^m)=p_i(x^n)p_i(y^m)=\prod_{t=1}^np_i(x_t)~\prod_{s=1}^mp_i(y_s),\nn
\ee
for all $n=1,2,\cdots$, $m=1,2,\cdots$. Then the optimal decision rules for the test with the Lagrangian (\ref{distr-lagrangian}) as objective are the following. At $X$, the decision rule is
\be
\label{distr-dec-rule-X}q_{\txt{opt}}(u_n=j|x^n)=I_{R_{u_n=j}}(x^n),
\ee
where the decision regions are
\be\ba
\label{distr-dec-regions-X}& R_{u_n=0}=\left\{x^n:{p_1(x^n)\over p_0(x^n)}<A^{(1)}\right\},\\
& R_{u_n=c}=\left\{x^n:A^{(1)}<{p_1(x^n)\over p_0(x^n)}<B^{(1)}\right\},\\
& R_{u_n=1}=\left\{x^n:B^{(1)}<{p_1(x^n)\over p_0(x^n)}\right\}.
\ea\ee
The thresholds $A,B$ are given by (\ref{d-thresholds-X}).

At $Y$, the decision rule is
\be
\label{distr-dec-rule-Y}q_{\txt{opt}}(v_m=j|y^m,u_n)=I_{R_{v_m=j|u_n}}(y^m),
\ee
where the decision regions are
\be\ba
\label{distr-dec-regions-Y}& R_{v_m=0|u_n}=\left\{y^m:{p_1(y^m)\over p_0(y^m)}<A^{(2)}_{u_n}\right\}\\
& R_{v_m=c|u_n}=\left\{y^m:A^{(2)}_{u_n}<{p_1(y^m)\over p_0(y^m)}<B^{(2)}_{u_n}\right\},\\
& R_{v_m=1|u_n}=\left\{y^m:B^{(2)}_{u_n}<{p_1(y^m)\over p_0(y^m)}\right\}.
\ea\ee
The thresholds $A,B$ are given by (\ref{d-thresholds-Y}).

Note that in (\ref{distr-dec-regions-X}),(\ref{d-thresholds-X}), we have suppressed dependence of the thresholds on $n$ for notational convenience. For the same reason, we have suppressed dependence on $m$ of the thresholds in (\ref{distr-dec-regions-Y}),(\ref{d-thresholds-Y}).
\end{thm}
\begin{IEEEproof}
See Appendix \ref{distr-thm-proof}.
\end{IEEEproof}

\begin{rmks}~

~~0) Without the conditional independence assumption in the theorem, the two-threshold likelihood rules are suboptimal in general.

~~1) Let $\tau_x,\tau_y$ denote the \emph{stopping times} as random variables. Then the \emph{stopping rules} are given by
\be\ba
&p(\tau_x=n|x^n) \triangleq q(u_n\neq c|x^n)\nn\\
&~~~~=1-q(u_n=c|x^n)=1-I_{R_{u_n=c}}(x^n),\nn\\
&p(\tau_y=m|y^m,u_n) \triangleq q(v_m\neq c|y^m,u_n)\nn\\
&~~~~=1-q(v_m=c|y^m,u_n)=1-I_{R_{v_m=c|u_n}}(y^m),\nn
\ea\ee
where $\tau_x\leq\tau_y$ due to the timing constraint $n\leq m$.

~~2) With the simplification (\ref{centr-grounding}) at $X$, the optimal value of the objective can be written as (with $\widehat{R}_{u_t=c}$, $\widehat{R}_{v_s=c}$, etc as in (\ref{centr-error-constraints-opt}))

\be\ba
\label{tandem-SO-opt}&S_{\txt{opt}}=\sum_{m,y^m,n\leq m,u^n,x^n,i}(m+1)\prod_{s=1}^mq_{\txt{opt}}(v_s=c|y^s,u_n)\\
&~~~~~~~~\times\prod_{t=1}^np_{\txt{opt}}(u_t|x^t,u^{t-1})~p_i(x^n,y^m)\pi_i\\
&~~~~=\sum_{m,y^m,n\leq m,u_n,x^n,i}(m+1)\prod_{s=1}^mq_{\txt{opt}}(v_s=c|y^s,u_n)\\
&~~~~~~~~\times q_{\txt{opt}}(u_n|x^n)\prod_{t=1}^{n-1}q_{\txt{opt}}(u_t=c|x^t)~p_i(x^n,y^m)\pi_i\\
&~~~~=\sum_{m,y^m,n,u_n,x^n,i}(m+1)\prod_{s=1}^mI_{R_{v_s=c|u_n}}(y^s)\\
&~~~~~~~~\times I_{R_{u_n}}(x^n)\prod_{t=1}^{n-1}I_{R_{u_t=c}}(x^t)~p_i(x^n,y^m)\pi_i\\
&~~~~=\sum_{m,y^m,n\leq m,u_n,x^n,i}(m+1)~I_{\cap_{s=1}^m\widehat{R}_{v_s=c|u_n}}(y^m)\\
&~~~~~~~~\times I_{R_{u_n}~\cap~\cap_{t=1}^{n-1}\widehat{R}_{u_t=c}}(x^n)~p_i(x^n,y^m)\pi_i\\
&~~~~\sr{CIO}{=}\sum_{m,y^m,n\leq m,u_n,x^n,i}(m+1)~I_{\cap_{s=1}^m\widehat{R}_{v_s=c|u_n}}(y^m)\\
&~~~~~~~~\times I_{R_{u_n}\cap~\cap_{t=1}^{n-1}\widehat{R}_{u_t=c}}(x^n)~p_i(x^n)p_i(y^m)\pi_i\\
&~~~~\sr{CIO}{=}\sum_{m,n\leq m,u_n,i}(m+1)~p_i\left(\mathop{\cap}\limits_{s=1}^m\widehat{R}_{v_s=c|u_n}\right)\\
&~~~~~~~~\times p_i\left(R_{u_n}\cap~\mathop{\cap}\limits_{t=1}^{n-1}\widehat{R}_{u_t=c}\right)\pi_i.
\ea\ee

3) The error constraints $p_0(v_m=1)\leq\al$, $p_1(v_m=0)\leq\beta$ can similarly be expressed as
\be\ba
\label{distr-error-constraints-opt}&\sum_{n\leq m,u_n}p_0\left(R_{v_m=1|u_n}\cap~\mathop{\cap}\limits_{s=1}^{m-1}\widehat{R}_{v_s=c|u_n}\right)\\
&~~~~~~~~\times p_0\left(R_{u_n}\cap~\mathop{\cap}\limits_{t=1}^{n-1}\widehat{R}_{u_t=c}\right)\leq\al,\\
&\sum_{n\leq m,u_n}p_1\left(R_{v_m=0|u_n}\cap~\mathop{\cap}\limits_{s=1}^{m-1}\widehat{R}_{v_s=c|u_n}\right)\\
&~~~~~~~~\times p_1\left(R_{u_n}\cap~\mathop{\cap}\limits_{t=1}^{n-1}\widehat{R}_{u_t=c}\right)\leq\beta,
\ea\ee
for $m=1,2,\cdots$
\end{rmks}

Note that unlike in the centralized test, the threshold values in $S_{\txt{opt}}$ are not completely determined by the error level constraints. Thus $S_{\txt{opt}}$ must be obtained through constrained minimization of $S$ as a function of the thresholds.

\subsection{Upper bound of the sample number}
The convergence result in Theorem \ref{cst-convergence} (along with the remarks following the theorem) extends to the distributed setting in a straightforward manner. The optimal decision rule for the sample-by-sample test is as follows: At $X$, we have
\be
\label{distr-dec-rule-X-sbs}q^{\txt{(sbs)}}(u_n=j|x_n)=I_{R^{\txt{(sbs)}}_{u_n=j}}(x_n),
\ee
where
\be\ba
\label{distr-dec-regions-X-sbs}& R^{\txt{(sbs)}}_{u_n=0}=\left\{x_n:{p_1(x_n)\over p_0(x_n)}<A^{(1)}\right\},\\
& R^{\txt{(sbs)}}_{u_n=c}=\left\{x_n:A^{(1)}<{p_1(x_n)\over p_0(x_n)}<B^{(1)}\right\},\\
& R^{\txt{(sbs)}}_{u_n=1}=\left\{x_n:B^{(1)}<{p_1(x_n)\over p_0(x_n)}\right\}.
\ea\ee
At $Y$, we have
\be
\label{distr-dec-rule-Y-sbs}q^{\txt{(sbs)}}(v_m=j|y_m,u_n)=I_{R^{\txt{(sbs)}}_{v_m=j|u_n}}(y_m),
\ee
where the decision regions are
\be\ba
\label{distr-dec-regions-Y-sbs}& R^{\txt{(sbs)}}_{v_m=0|u_n}=\left\{y_m:{p_1(y_m)\over p_0(y_m)}<A^{(2)}_{u_n}\right\}\\
& R^{\txt{(sbs)}}_{v_m=c|u_n}=\left\{y_m:A^{(2)}_{u_n}<{p_1(y_m)\over p_0(y_m)}<B^{(2)}_{u_n}\right\},\\
& R^{\txt{(sbs)}}_{v_m=1|u_n}=\left\{y_m:B^{(2)}_{u_n}<{p_1(y_m)\over p_0(y_m)}\right\}.
\ea\ee
Thus, we have a sample-by-sample processing bound
\be\ba
&S_{\txt{opt}}=\sum_{m,n\leq m,u_n,i}(m+1)~p_i\left(\mathop{\cap}\limits_{s=1}^m\widehat{R}_{v_s=c|u_n}\right)\\
&~~~~~~~~\times p_i\left(R_{u_n}\cap~\mathop{\cap}\limits_{t=1}^{n-1}\widehat{R}_{u_t=c}\right)\pi_i\\
&~~~~\leq\sum_{m,n\leq m,u_n,i}(m+1)~p_i\left(\prod_{s=1}^mR^{(\txt{sbs})}_{v_s=c|u_n}\right)\\
&~~~~~~~~\times p_i\left(R_{u_n}^{(\txt{sbs})}\times\prod_{t=1}^{n-1}R^{(\txt{sbs})}_{u_t=c}\right)\pi_i\\
&~~~~=\sum_{m,n\leq m,u_n,i}(m+1)~p_i\left(\prod_{s=1}^mR^{(\txt{sbs})}_{v_s=c|u_n}\right)\\
&~~~~~~~~\times p_i\left(R_{u_n}^{(\txt{sbs})}\right)p_i\left(\prod_{t=1}^{n-1}R^{(\txt{sbs})}_{u_t=c}\right)\pi_i\\
\label{distr-SBS-CIO-bound}&~~~~\leq\sum_{m,n\leq m,u,i} (m+1)~e^{-mL^{(2)}_{i,u}}~e^{-L^{(1)}_i(u)}e^{-(n-1)L_i^{(1)}(c)}~\pi_i\\
&~~~~=\sum_{m,u,i} (m+1)~e^{-mL^{(2)}_{i,u}}~{(1-e^{-mL^{(1)}_i(c)})\over 1-e^{-L^{(1)}_i(c)}}e^{-L^{(1)}_i(u)}~\pi_i\\
&~~~~=\sum_{u,i}(S_{i,u}-T_{i,u}){e^{-L^{(1)}_i(u)}\over 1-e^{-L^{(1)}_i(c)}}~\pi_i,\\
\ea\ee
where
\be\ba
&S_{i,u}=\sum_m(m+1)~e^{-mL^{(2)}_{i,u}}={2~\!e^{L^{(2)}_{i,u}}-1\over\left(e^{L^{(2)}_{i,u}}-1\right)^2},\nn\\
&T_{i,u}=\sum_m(m+1)~e^{-m\left[L^{(2)}_{i,u}+L^{(1)}_i(c)\right]},\nn\\
&~~~~={2~\!e^{L^{(2)}_{i,u}+L^{(1)}_i(c)}-1\over\left(e^{L^{(2)}_{i,u}+L^{(1)}_i(c)}-1\right)^2},\nn
\ea\ee
and

{\small
\be\ba
&L_i^{(1)}(u)=\left\{
                  \begin{array}{ll}
                   -\log~p_i\big(p_1(x)/p_0(x)<A^{(1)}\big), & u=0 \\
                   -\log~p_i\big(A^{(1)}<p_1(x)/p_0(x)<B^{(1)}\big), & u=c\\
                   -\log~p_i\big(B^{(1)}<p_1(x)/p_0(x)\big), & u=1
                  \end{array}
                \right.\nn\\
&L^{(2)}_{i,u}=-\log~p_i\big(A^{(2)}_{u}<p_1(x)/p_0(x)<B^{(2)}_{u}\big).\nn
\ea\ee
}

In analogy with (\ref{distr-error-constraints-opt}), the constraints for the sample-by-sample objective in (\ref{distr-SBS-CIO-bound}) are
\be\ba
\label{distr-error-constraints-sbs}&\sum_{n\leq m,u_n}p_0\left(R^{\txt{(sbs)}}_{v_m=1|u_n}\times\prod_{s=1}^{m-1}R^{\txt{(sbs)}}_{v_s=c|u_n}\right)\\
&~~~~~~~~\times p_0\left(R^{\txt{(sbs)}}_{u_n}\times\prod_{t=1}^{n-1}R^{\txt{(sbs)}}_{u_t=c}\right)\leq\al,\\
&\sum_{n\leq m,u_n}p_1\left(R^{\txt{(sbs)}}_{v_m=0|u_n}\times\prod_{s=1}^{m-1}R^{\txt{(sbs)}}_{v_s=c|u_n}\right)\\
&~~~~~~~~\times p_1\left(R^{\txt{(sbs)}}_{u_n}\times\prod_{t=1}^{n-1}R^{\txt{(sbs)}}_{u_t=c}\right)\leq\beta,
\ea\ee
for $m=1,2,\cdots$, which can be expressed as
\be\ba
\label{distr-error-constraints-sbs2}& \max_m~\sum_ue^{-mL^{(2)}_{0,u}}~{(1-e^{-mL^{(1)}_i(c)})\over 1-e^{-L^{(1)}_0(c)}}e^{-L^{(1)}_0(u)}\leq \al,\\
& \max_m~\sum_ue^{-mL^{(2)}_{1,u}}~{(1-e^{-mL^{(1)}_1(c)})\over 1-e^{-L^{(1)}_1(c)}}e^{-L^{(1)}_1(u)}\leq \beta.
\ea\ee

For the purpose of numerical computation and comparison with the centralized tests, it will suffice to consider the following special case. Let us assume $X$ has already stopped sampling at some fixed $n\leq S_{\txt{opt}}$, and let
\be\ba
&S_n=\sum_{m\geq n,u_n,i}(m+1)~p_i\left(\mathop{\cap}\limits_{s=1}^m\widehat{R}_{v_s=c|u_n}\right)\\
&~~~~~~~~\times p_i\left(R_{u_n}\cap~\mathop{\cap}\limits_{t=1}^{n-1}\widehat{R}_{u_t=c}\right)\pi_i.\nn
\ea\ee
Since $1\leq S_{\txt{opt}}$ is satisfied, it follows that such an $n$ is possible. In particular, with $n=1$ (i.e., X makes only one observation), we get
\be\ba
&S_1=\sum_{m,u_1,i}(m+1)~p_i\left(\mathop{\cap}\limits_{s=1}^m\widehat{R}_{v_s=c|u_1}\right)~ p_i\left(R_{u_1}\right)\pi_i\\
&~~~~\leq \sum_{m,u_1,i}(m+1)~p_i\left(\prod_{s=1}^mR^{(\txt{sbs})}_{v_s=c|u_1}\right)~ p_i\left(R^{(\txt{sbs})}_{u_1}\right)\pi_i\\
\label{distr-SBS-CIO-bound2}&~~~\leq\sum_{m,u_,i} (m+1)~e^{-mL^{(2)}_{i,u_1}}~e^{-L^{(1)}_i(u_1)}~\pi_i\\
&~~~~=\sum_{u_1,i} {2~\!e^{L^{(2)}_{i,u_1}}-1\over\left(e^{L^{(2)}_{i,u_1}}-1\right)^2}~e^{-L^{(1)}_i(u_1)}~\pi_i\\
\ea\ee
where
\be\ba
&L^{(2)}_{i,u_1}=-\log~p_i\big(A^{(2)}_{u_1}<p_1(x)/p_0(x)<B^{(2)}_{u_1}\big),\nn
\ea\ee
and since X makes only one observation, we have
{\small
\be\ba
& e^{-L^{(1)}_i(u_1)}=p_i\left(R^{(\txt{sbs})}_{u_1}\right)=p_i\left(R_{u_1}\right)\\
& ~~~~=\left\{
                  \begin{array}{ll}
                   p_i\big(p_1(x)/p_0(x)<A^{(1)}\big), & u_1=0 \\
                   p_i\big(A^{(1)}<p_1(x)/p_0(x)<B^{(1)}\big), & u_1=c\\
                   p_i\big(B^{(1)}<p_1(x)/p_0(x)\big), & u_1=1
                  \end{array}
                \right.\nn
\ea\ee
}Working with Example \ref{S-test-example}, the relevant probabilities in terms of $Q$-functions are
\be\ba
& p_i\big(B<p_1(x)/p_0(x)\big)=Q\big((b-i)/\sigma\big),\nn\\
& p_i\big(A<p_1(x)/p_0(x)<B\big)\nn\\
&~~~~~~=Q\big((a-i)/\sigma\big)-Q\big((b-i)/\sigma\big),\nn\\
& p_i\big(p_1(x)/p_0(x)<A\big)=1-Q\big((a-i)/\sigma\big),\nn\\
& a=\sigma^2\log A+1/2,~~~~b=\sigma^2\log B+1/2,\nn
\ea\ee
where $A$ is $A^{(1)}$ or $A^{(2)}_{u_1}$ and $B$ is $B^{(1)}$ or $B^{(2)}_{u_1}$.

The bound in (\ref{distr-SBS-CIO-bound2}) gives
\be\ba
\label{distr-SBS-CIO-bound2.1}S_1\leq \sum_{u_1,i}f(\Delta_{i,u_1})P_i(u_1)\pi_i,
\ea\ee
where~ $\Delta_{i,u_1}=Q\big((a^{(2)}_{u_1}-i)/\sigma_2\big)-Q\big((b^{(2)}_{u_1}-i)/\sigma_2\big)$,

{\small
\be\ba
&P_i(u_1)=\left\{
                  \begin{array}{ll}
                   1-Q\left((a^{(1)}-i)/\sigma_1\right), & u_1=0 \\
                   Q\left((a^{(1)}-i)/\sigma_1\right)-Q\left((b^{(1)}-i)/\sigma_1\right), & u_1=c\\
                   Q\left((b^{(1)}-i)/\sigma_1\right), & u_1=1
                  \end{array}
                \right.\nn
\ea\ee
}and $f$ is the function in (\ref{base-function}).

\begin{figure}[H]
\centering
\scalebox{0.6}{\includegraphics[width=15cm,height=9cm]{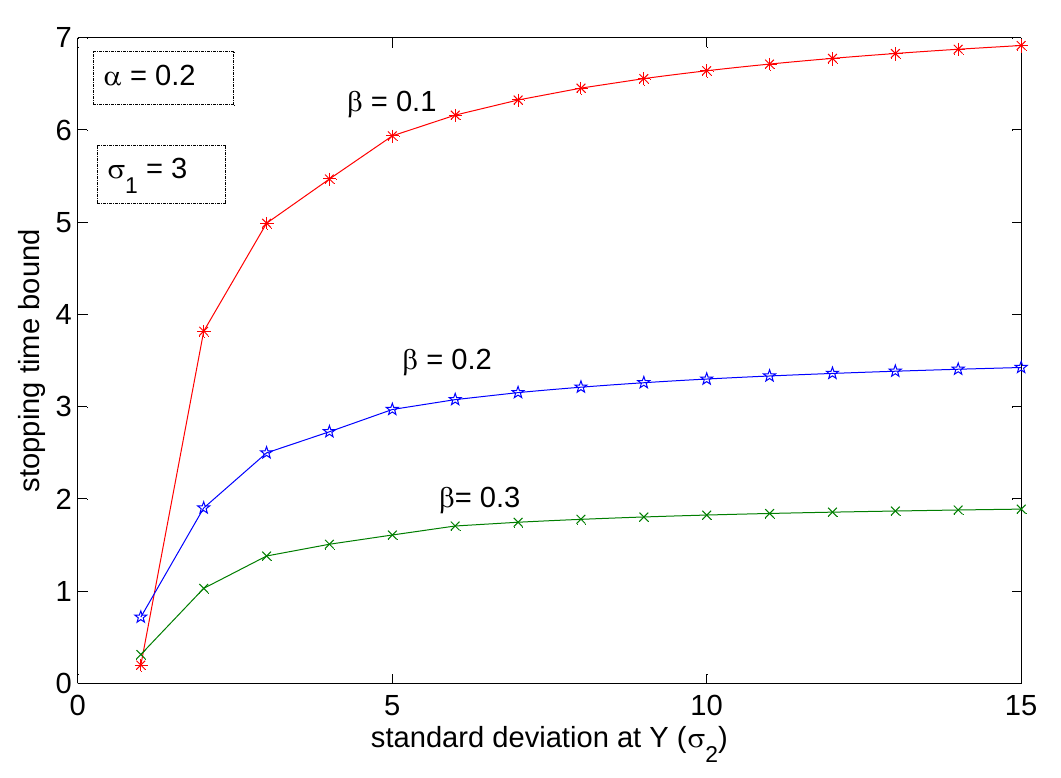}}\\
\caption{Two-sensor distributed test: Dependence of the stopping time bound on standard deviation at Y and missed detection level $\beta$.}\label{Distributed-SN}
\end{figure}

As distributed versions of the sample-by-sample constraints (\ref{centr-SBS-constraints}), we have the constraints
\be\ba
\label{distr-rect-constraints}&\sum_{u_1}p_0\big(B^{(2)}_{u_1}<p_1(x)/p_0(x)\big)e^{-L^{(1)}_0(u_1)}\\
&~~~~=\sum_{u_1}Q\big(b^{(2)}_{u_1}/\sigma_2\big)P_0(u_1)\leq\al,\\
&\sum_{u_1}p_1\big(p_1(x)/p_0(x)<A^{(2)}_{u_1}\big)e^{-L^{(1)}_1(u_1)}\\
&~~~~=\sum_{u_1}\left(1-Q\big((a^{(2)}_{u_1}-1)/\sigma_2\big)\right)P_1(u_1)\leq\beta.
\ea\ee
We need to minimize the bound (\ref{distr-SBS-CIO-bound2.1}) over the eight threshold parameters $a^{(1)},b^{(1)},a^{(2)}_{u_1},b^{(2)}_{u_1}$ subject to the constraints (\ref{distr-rect-constraints}). The numerical results are shown in Fig \ref{Distributed-SN}.

\begin{figure}[H]
\centering
\scalebox{0.6}{\includegraphics[width=15cm,height=9cm]{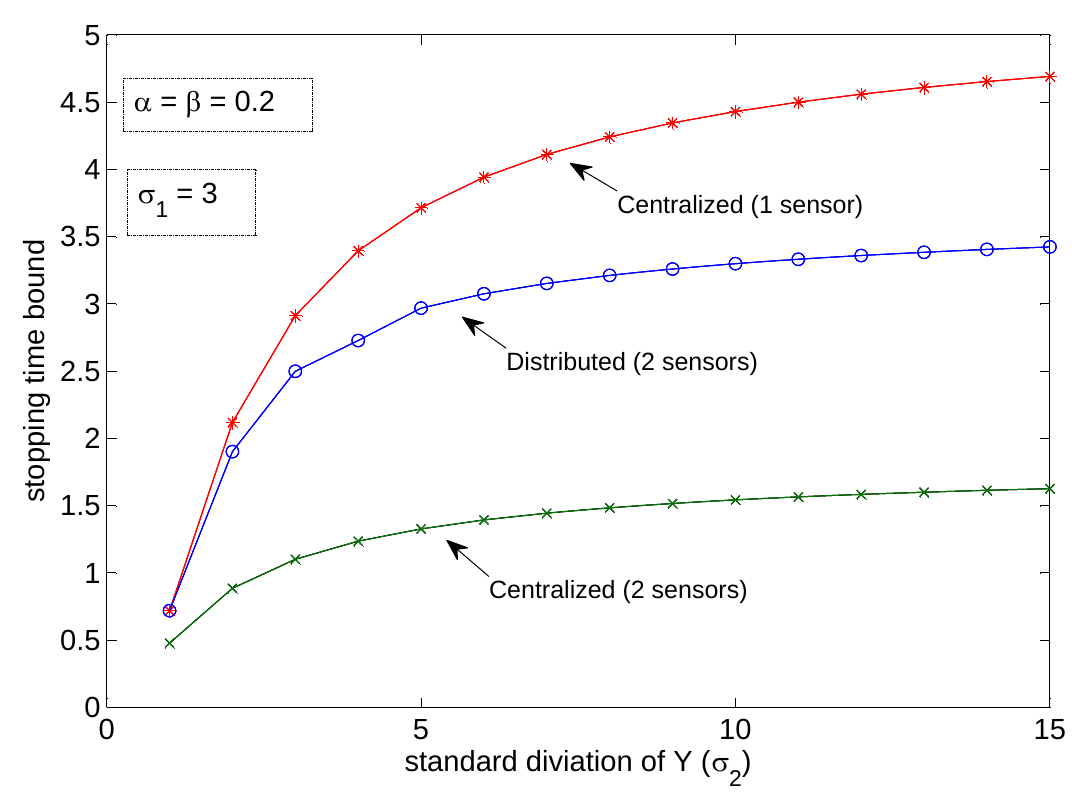}}\\
\caption{Relative performance of the upper bound for (1) the one-sensor centralized test, (2) the two-sensor distributed test, and (3) the two-sensor centralized test.}\label{Combined-SN}
\end{figure}

Fig \ref{Combined-SN} compares performance of the upper bound for the three different situations we have studied, namely, $Y$ as a one-sensor network, $\{X,Y\}$ as a centralized two-sensor network, and $\{X,Y\}$ as a distributed two-sensor network. As indicated in the figure, we fix the standard deviation at $\sigma_1=3$ and the detection error levels at $\al=\beta=0.2$. It is evident from the graphs that distributed processing is most beneficial when observations have low quality.

From Figures \ref{Centralized-SN}, \ref{2Sensor-Centralized-SN}, \ref{Distributed-SN}, \ref{Combined-SN}, we can deduce that the optimal stopping time has a finite upper bound, the behavior of which is precisely what one expects of the behavior of the true stopping time. For sufficiently simple networks in particular, the behavior of the optimal stopping time may not differ qualitatively from that of its upper bound in any interesting way.  

\section{Generalization: Acyclic directed graphs}\label{generalize}
In this section, we extend the results of the previous sections to any detection network in the form of an acyclic directed graph. In order to establish notation, we will begin by providing the usual binary decision rules for static detection over acyclic graph networks in Section \ref{generalize-ss1}. This is then followed by a discussion of the sequential test for such networks in Section \ref{generalize-ss2}.

\subsection{Binary decision rules for static detection over acyclic graphs}\label{generalize-ss1}
Let $\vec{X}=\{X_1,...,X_K\}$ be a network of $K$ sensors. Without loss of generality, we assume $X_1$ is the fusion center. Let $\td{X}_k=\{X_{k_1},X_{k_2},...,X_{k_{I_k}}\}\subset \vec{X}$ denote the parents of $X_k$, i.e., all sensors forwarding their decisions to $X_k$.

Denoting the observation of $X_k$ by $x_k$ and its decision by $u_k$, the dependence structure of the decision is given by
\bea
u_k=\gamma_k(x_k,\td{u}_k),\nn
\eea
where $\td{u}_k=\{u_{k_1},u_{k_2},...,u_{k_{I_k}}\}$ consists of the decisions of parents of $X_k$. Let $\vec{x}=(x_1,...,x_K)$ and $\vec{u}=(u_1,...,u_K)$, and consider the risk function
\be\ba
\label{acyclic-bayes-objective}& S=\sum_{u_1,i}C_{u_1i}p(u_1,i)=\sum_{\vec{x},\vec{u},i}C_{u_1i}~\prod_{k=1}^Kp(u_k|x_k,\td{u}_k)~p_i(\vec{x})\pi_i\\
&~~~~\sr{(a)}{=}\sum_{\vec{x},\vec{u},i}C_{u_1i}~\prod_{k=1}^Kp(u_k|x_k,\td{u}_k)~\prod_{k=1}^Kp_i(x_k)\pi_i,
\ea\ee
where step (a) holds for conditionally independent observations.

For simplicity, we require that every sensor sends the same message to all of its offsprings in the network.

\begin{thm}\label{acyclic-theorem}
Based on the objective function (\ref{acyclic-bayes-objective}), and assuming observations are conditionally independent, the optimal binary decision rule for the network $\{X_1,...,X_K\}$ viewed as an acyclic directed graph is
\be
\label{acyclic-bayes-rule}p_{\txt{opt}}(u_k=1|x_k,\td{u}_k)=I_{R_{u_k=1|\td{u}_k}}(x_k),
\ee
where the decision regions have the form
\be\ba
\label{acyclic-bayes-region}R_{u_{k}=1|\td{u}_{k}}=\left\{x_{k}:{p_1(x_{k})\over p_0(x_{k})}>\ld^{(k)}_{\td{u}_{k}}\right\}.
\ea\ee

The optimal value of the risk function is
\be\ba
\label{acyclic-graph-riskeq}& S_{\txt{opt}}=\sum_{\vec{x},\vec{u},i}C_{u_1i}~\prod_{k=1}^KI_{R_{u_k|\td{u}_k}}(x_k)~\prod_{k=1}^Kp_i(x_k)~\pi_i\\
&~~\sr{(a)}{=}\sum_{\vec{u},i}C_{u_1i}~\prod_{k=1}^Kp_i(R_{u_k|\td{u}_k})~\pi_i,
\ea\ee
where step (a) holds due to conditionally independent observations.
\end{thm}
\begin{IEEEproof}
This is the result of Theorem \ref{acyclic-prop} simplified for binary decisions.
\end{IEEEproof}

\subsection{The sequential test for acyclic directed graphs}\label{generalize-ss2}
Consider the network of $K$ sensors $X=\{X_k:k=1,...,K\}=(X_1,...,X_K)$, with $\td{X}_k=(X_{k_1},...,X_{k_{I_k}})\subset X$ denoting the parents of $X_k$. If $X_k$ makes a sequence of $n_k$ observations, where $n_k\in\{1,2,\cdots\}$, we denote these observations by $x_k^{n_k}=(x_{k1},x_{k2},\cdots, x_{kn_k})$. Similarly, the decisions $u_k^{n_k}=(u_{k1},u_{k2},...,u_{kn_k})$ of $X_k$ are given by
\be
u_{kn_k}=\gamma_{kn_k}\left(x_{k}^{n_k},u_k^{n_k-1},\td{u}_{k\td{n}_k}\right),
\ee
where $\td{u}_{k\td{n}_k}=(u_{k_1n_{k_1}},u_{k_2n_{k_2}},\cdots,u_{k_{I_k}n_{k_{I_k}}})$ is the set of decisions forwarded to $X_k$, and we impose the timing constraints
\be
n_{k_j}\leq n_k,~~~~j=1,...,I_k.\nn
\ee
That is, we assume that $X_k$ begins sampling only after receiving the decisions of all of its parents. For notational convenience, $\sum_{\vec{n},\{\max(\td{n}_k)\leq n_k\}}$ will be written simply as $\sum_{\vec{n}}$ with the timing constraints understood.

For simplicity, we require that every sensor sends the same message to all of its offsprings in the network. We will write
\be\ba
& \vec{n}=(n_1,...,n_k),~~\vec{x}_{\vec{n}}=(x_{1n_1},...,x_{kn_k}),\nn\\
&\vec{x}^{\vec{n}}=(x_1^{n_1},x_2^{n_2},...,x_K^{n_K}),~~\vec{u}^{\vec{n}}=(u_1^{n_1},u_2^{n_2},...,u_K^{n_K}),\\
&q\left(u_{kn_k}=j|x_k^{n_k},\td{u}_{k\td{n}_k}\right)=p\left(u_{kn_k}=j|x_k^{n_k},u^{n_k-1}_k=c,\td{u}_{k\td{n}_k}\right).\nn
\ea\ee

We consider \emph{two} hypotheses
\bea
H_i:(\vec{x}^{\vec{n}})\sim p_i(\vec{x}^{\vec{n}})=p(\vec{x}^{\vec{n}}|H_i),~~i=0,1,
\eea
and \emph{ternary} decision rules
\be\ba
&\gamma_{kn_k}:(x_k^{n_k},u_k^{n_k-1},\td{u}_{k\td{n}_k})\in\X_k^{n_k}\times\U_k^{n_k-1}\times\td{\U}_{k\td{n}_k}\\
&~~~~\longmapsto u_{kn_k}=\gamma_{kn_k}(x_k^{n_k},u_k^{n_k-1},\td{u}_{k\td{n}_k})\in\U_{kn_k}=\{0,c,1\}.\nn
\ea\ee
Here, $u_{kn_k}=0$ is acceptance of $H_0$, $u_{kn_k}=1$ is acceptance of $H_1$, and $u_{kn_k}=c$ is rejection of both $H_0$ and $H_1$ in favor of continued sampling.
At $X_k$, for any given $\td{n}_k$, the next sample $x_{k,n_k+1}$ is taken if and only if
\be
\gamma_{kt_k}(x_k^{t_k},u_k^{t_k-1},\td{u}_{k\td{t}_k})=c~~\txt{for all}~~\max(\td{n}_k)\leq t_k\leq n_k.
\ee

Therefore, for each $\td{n}_k=1,2,\cdots$, the sample size $n_k=h_{n_k}(u_k^{n_k-1},\td{u}_{k\td{n}_k})$ is a random variable whose distribution $p(n_k)$ satisfies
\be\ba
\label{distr-stop-time-acyclic}&p(n_k+1)=p(u_k^{n_k}=c)\triangleq p(u_{k1}=c,...,v_{kn_k}=c),\\
&\sum_{n_k=1}^\infty p(u_k^{n_k}=c)=1,\\
&p(u_{kn_k}=j|u_{kt_k}=j'\neq c)=\delta_{jj'},~~\max(\td{n}_k)\leq t_k<n_k.
\ea\ee

At the fusion center $X_1$, given a \emph{false alarm} level $p_0(u_{n_1}=1)=p(u_{n_1}=1|H_0)\leq\al$ and a \emph{missed detection} level $p_1(u_{n_1}=0)=p(u_{n_1}=0|H_1)\leq\beta$, our objective is to minimize the expected number of samples
\be\ba
\label{distr-sequential-objective-acyclic}&S=\sum_{n_1=1}^\infty(n_1+1)p(n_1+1)=\sum_{n_1=1}^\infty (n_1+1)p(u_1^{n_1}=c)\\
&~~=\sum_{\vec{n},\vec{u}^{\hat{n}_1},\vec{x}^{\vec{n}},i}(n_1+1)p(u_1^{n_1}=c,\vec{u}^{\hat{n}_1},\hat{n}_1,\vec{x}^{\vec{n}},H_i)\\
&~~=\sum_{\vec{n},\vec{u}^{\hat{n}_1},\vec{x}^{\vec{n}},i}(n_1+1)\pi_ip_i(\vec{x}^{\vec{n}})~p(u_1^{n_1}=c|x_1^{n_1},\td{u}_{1\td{n}_1})\\
&~~~~\times p(\vec{u}^{\hat{n}_1}|\vec{x}^{\hat{n}_1})\\
&~~=\sum_{\vec{n},\vec{u}_{\hat{n}_1},\vec{x}^{\vec{n}},i}(n_1+1)\pi_ip_i(\vec{x}^{\vec{n}})~\prod_{t_1=1}^{n_1}q(u_{1t_1}=c|x_1^{t_1},\td{u}_{1\td{t}_1})\\
&~~\times\prod_{k=2}^K\left(q(u_{kn_k}|x_k^{n_k},\td{u}_{k\td{n}_k})\prod_{t_k=1}^{n_k-1}q(u_{kt_k}=c|x_k^{t_k},\td{u}_{k\td{t}_k})\right)\\
&~~=\sum_{n_1=1}^\infty (n_1+1)p(u_{1n_1}=c),
\ea\ee
where $\hat{n}_k=\vec{n}\backslash n_k$, $\vec{u}^{\hat{n}_k}=\vec{u}^{\vec{n}}\backslash u_k^{n_k}$, and
\be
\vec{u}_{\hat{n}_k}=\vec{u}_{\vec{n}}\backslash u_{kn_k}.\nn
\ee

Thus, we have the optimization problem
\be\ba
&\txt{minimize}~S\\
&\txt{subject to}~\{p_0(u_{1n_1}=1)\leq\al,~p_1(u_{1n_1}=0)\leq\beta\}_{n_1},\\
&~~~~n_1=1,2,...,~~~~\sum_{n_1=1}^\infty p(u_1^{n_1}=c)=1,
\ea\ee
where the optimization variables are
\bea
\{q(u_{kn_k}=j|x_k^{n_k},\td{u}_{k\td{n}_k}),\}_{j,\vec{n},\vec{x}^{\vec{n}},\vec{\td{u}}_{\vec{\td{n}}}}.\nn
\eea

The Lagrangian for the above problem is
\be\ba
&L=\sum_{n_1}\bigg((n_1+1)p(u_1^{n_1}=c)+\ld_{n_10}\big[\al-p_0(u_{1n_1}=1)\big]\\
&~~~~+\ld_{n_11}\big[\beta-p_1(u_{1n_1}=0)\big]+\ld\big[s_{n_1}-p(u_1^{n_1}=c)\big]\bigg),\nn
\ea\ee
where~ $\ld_{n_10},\ld_{n_11}\leq 0,\ld\in\Real$,~ $p(u_1^{n_1}=c)=p_0(u_1^{n_1}=c)\pi_0+p_1(u_1^{n_1}=c)\pi_1$,~ and $s_n$ is any sequence of numbers such that $\sum s_n=1$. We can perform the rescaling $\ld_{n_1i}\ra(n_1+1)\ld_{n_1i}$, $\ld\ra (n_1+1)\ld$ without loss of generality. Thus we can write
\be\ba
\label{distr-lagrangian-acyclic}&L=\sum_{n_1}(n_1+1)\bigg(p(u_1^{n_1}=c)+\ld_{n_10}\big[\al-p_0(u_{1n_1}=1)\big]\\
&~~~~+\ld_{n_11}\big[\beta-p_1(u_{1n_1}=0)\big]+\ld\big[s_{n_1}-p(u_1^{n_1}=c)\big]\bigg),\\
&p(u_1^{n_1}=c)=p_0(u_1^{n_1}=c)\pi_0+p_1(u_1^{n_1}=c)\pi_1,
\ea\ee
where we have the expansion
\be\ba
\label{distr-terms-expansion-acyclic}&p_i(u_{1n_1}=j)\\
&~~=\sum_{\hat{n}_1,u_1^{n_1-1},\vec{u}^{\hat{n}_1},\vec{x}^{\vec{n}}}p_i(u_{1n_1}=j,u_1^{n_1-1},\vec{u}^{\hat{n}_1},\hat{n}_1,\vec{x}^{\vec{n}})\\
&~~=\sum p(u_{1n_1}=j,u_1^{n_1-1}|x_1^{n_1},\td{u}_{1\td{n}_1})\\
&~~~~\times p(\bar{u}^{\hat{n}_1}|\bar{x}^{\hat{n}_1})p_i(\vec{x}^{\vec{n}})\\
&~~=\sum_{\hat{n}_1,\vec{u}_{\hat{n}_1},\vec{x}^{\vec{n}}} p_i(\vec{x}^{\vec{n}})\\
&~~\times q(u_{1n_1}=j|x_1^{n_1},\td{u}_{1\td{n}_1})\prod_{t_1=1}^{n_1-1}q(u_{1t_1}=c|x_1^{t_1},\td{u}_{1\td{t}_1})\\
&~~\times\prod_{k=2}^K\left(q(u_{kn_k}|x_k^{n_k},\td{u}_{k\td{n}_k})\prod_{t_k=1}^{n_k-1}q(u_{kt_k}=c|x_k^{t_k},\td{u}_{k\td{t}_k})\right)\\
&~~=p_i(u_{1n_1}=j,u_1^{n_1-1}=c).
\ea\ee

The following result is a straightforward extension of Theorem \ref{distr-thm}.
\begin{thm}\label{distr-thm-acyclic}
Suppose the observation samples $\vec{x}^{\vec{n}}=(x_1^{n_1},...,x_K^{n_K})$ of the acyclic directed graph network $(X_1,...,X_K)$ are conditionally independent so that
\be
p_i(\vec{x}^{\vec{n}})=\prod_{k=1}^Kp_i(x_k^{n_k})=\prod_{k=1}^K~\prod_{t_k=1}^{n_k}p_i(x_{kt_k}),\nn
\ee
for all $n_k=1,2,\cdots,$ $k=1,\cdots,K$. Then we have the following optimal decision strategy for the test with the Lagrangian (\ref{distr-lagrangian-acyclic}) as objective. At $X_k$, the decision rule is
\be
\label{distr-dec-rule-acyclic}q_{\txt{opt}}\left(u_{kn_k}=j|x_k^{n_k},\td{u}_{k\td{n}_k}\right)=I_{R_{u_{kn_k}=j|\td{u}_{k\td{n}_k}}}\left(x_k^{n_k}\right),
\ee
where $j=0,c,1$, and the decision regions are
\be\ba
\label{distr-dec-regions-acyclic}&R_{u_{kn_k}=0|\td{u}_{k\td{n}_k}}=\left\{x_k^{n_k}:{p_1\left(x_k^{n_k}\right)\over p_0\left(x_k^{n_k}\right)}<A^{(k)}_{n_k,\td{u}_{k\td{n}_k}}\right\},\\
&R_{u_{kn_k}=c|\td{u}_{k\td{n}_k}}=\left\{x_k^{n_k}:A^{(k)}_{n_k,\td{u}_{k\td{n}_k}}<{p_1\left(x_k^{n_k}\right)\over p_0\left(x_k^{n_k}\right)}<B^{(k)}_{n_k,\td{u}_{k\td{n}_k}}\right\},\\
&R_{u_{kn_k}=1|\td{u}_{k\td{n}_k}}=\left\{x_k^{n_k}:B^{(k)}_{n_k,\td{u}_{k\td{n}_k}}<{p_1\left(x_k^{n_k}\right)\over p_0\left(x_k^{n_k}\right)}\right\}.
\ea\ee
\end{thm}
\begin{IEEEproof}
See the proof of Theorem (\ref{distr-thm}).
\end{IEEEproof}

\begin{rmks}~

~~0) Without the conditional independence assumption in the theorem, the two-threshold likelihood rules are suboptimal in general.

~~1) Let $\tau_k$ denote the \emph{stopping time} of $X_k$ as a random variable. Then the \emph{stopping rule} at $X_k$ is given by
\be\ba
&p(\tau_k=n_k|x_k^{n_k},\td{u}_{k\td{n}_k}) \triangleq q(u_{kn_k}\neq c|x_k^{n_k},\td{u}_{k\td{n}_k})\nn\\
&~~~~=1-q(u_{kn_k}=c|x_k^{n_k},\td{u}_{k\td{n}_k})=1-I_{R_{u_{kn_k}=c|\td{u}_{k\td{n}_k}}}(x_k^{n_k}),\nn
\ea\ee
where $\max(\td{\tau}_k)\leq\tau_k$ due to the timing constraint $\max(\td{n}_k)\leq n_k$.

~~2) In terms of the decision regions, the optimal value of the objective can be written as

\be\ba
\label{distr-sequential-objective-acyclic-opt}&S_{\txt{opt}}=\sum_{\vec{n},\vec{x}^{\vec{n}},\vec{u}^{\hat{n}_1},i} (n_1+1)\pi_ip_i(\vec{x}^{\vec{n}})~\prod_{t_1=1}^{n_1}q_{\txt{opt}}(u_{1t_1}=c|x_1^{t_1},\td{u}_{1\td{t}_1})\\
&~~\times\prod_{k=2}^K\left(q(u_{kn_k}|x_k^{n_k},\td{u}_{k\td{n}_k})\prod_{t_k=1}^{n_k-1}q_{\txt{opt}}(u_{kt_k}=c|x_k^{t_k},\td{u}_{k\td{t}_k})\right)\\
&~~~~=\sum_{\vec{n},\vec{u}_{\hat{n}_1},i} (n_1+1)\pi_i~p_i\left(\mathop{\cap}\limits_{t_1=1}^{n_1}\widehat{R}_{u_{1t_1}=c|\td{u}_{1\td{t}_1}}\right)\\
&~~\times\prod_{k=2}^Kp_i\left(R_{u_{kn_k}|\td{u}_{k\td{n}_k}}\cap~\mathop{\cap}\limits_{t_k=1}^{n_k-1}\widehat{R}_{u_{kt_k}=c|\td{u}_{k\td{t}_k}}\right),
\ea\ee
where $\widehat{R}_{u_{kt_k}=c|\td{u}_{k\td{t}_k}}=\Real^{n_k-t_k}\times R_{u_{kt_k}=c|\td{u}_{k\td{t}_k}}$.

3) The optimal value of the probability (\ref{distr-terms-expansion-acyclic}) is
\be\ba
\label{distr-terms-expansion-acyclic-opt}&p_i(u_{1n_1}=j)\\
&~~=\sum_{\hat{n}_1,u_1^{n_1-1},\vec{u}^{\hat{n}_1},\vec{x}^{\vec{n}}} p_i(\vec{x}^{\vec{n}})\\
&~~\times q_{\txt{opt}}(u_{1n_1}=j|x_1^{n_1},\td{u}_{1\td{n}_1})\prod_{t_1=1}^{n_1-1}q_{\txt{opt}}(u_{1t_1}=c|x_1^{t_1},\td{u}_{1\td{t}_1})\\
&~~\times\prod_{k=2}^K\left(q_{\txt{opt}}(u_{kn_k}|x_k^{n_k},\td{u}_{k\td{n}_k})\prod_{t_k=1}^{n_k-1}q_{\txt{opt}}(u_{kt_k}=c|x_k^{t_k},\td{u}_{k\td{t}_k})\right)\\
&~~=\sum_{\hat{n}_1,\vec{u}_{\hat{n}_1}} p_i\left(R_{u_{1n_1}=j|\td{u}_{1\td{n}_1}}\cap~\mathop{\cap}_{t_1=1}^{n_1-1}\widehat{R}_{u_{1t_1}=c|\td{u}_{1\td{t}_1}}\right)\\
&~~\times\prod_{k=2}^Kp\left(R_{u_{kn_k}|\td{u}_{k\td{n}_k}}\cap~\mathop{\cap}_{t_k=1}^{n_k-1}\widehat{R}_{u_{kt_k}=c|\td{u}_{k\td{t}_k}}\right),
\ea\ee
which can be used to express the error constraints $p_0(u_{1n_1}=1)\leq\al$, $p_1(u_{1n_1}=0)\leq\beta$.
\end{rmks}

\subsection{Upper bound of the sample number}
The decision rule for the sample-by-sample test is
\be
\label{distr-dec-rule-acyclic-sbs}q^{\txt{(sbs)}}\left(u_{kn_k}=j|x_{kn_k},\td{u}_{k\td{n}_k}\right)=I_{R^{\txt{(sbs)}}_{u_{kn_k}=j|\td{u}_{k\td{n}_k}}}\left(x_{kn_k}\right),
\ee
where the decision regions are
\be\ba
\label{distr-dec-regions-acyclic}&R^{\txt{(sbs)}}_{u_{kn_k}=0|\td{u}_{k\td{n}_k}}=\left\{x_{kn_k}:{p_1\left(x_{kn_k}\right)\over p_0\left(x_{kn_k}\right)}<A^{(k)}_{n_k,\td{u}_{k\td{n}_k}}\right\}\\
&R^{\txt{(sbs)}}_{u_{kn_k}=c|\td{u}_{k\td{n}_k}}=\left\{x_{kn_k}:A^{(k)}_{n_k,\td{u}_{k\td{n}_k}}<{p_1\left(x_{kn_k}\right)\over p_0\left(x_{kn_k}\right)}<B^{(k)}_{n_k,\td{u}_{k\td{n}_k}}\right\}\\
&R^{\txt{(sbs)}}_{u_{kn_k}=1|\td{u}_{k\td{n}_k}}=\left\{x_{kn_k}:B^{(k)}_{n_k,\td{u}_{k\td{n}_k}}<{p_1\left(x_{kn_k}\right)\over p_0\left(x_{kn_k}\right)}\right\}.
\ea\ee

Thus, as before, we have a sample-by-sample processing bound
\be\ba
\label{distr-SBS-CIO-bound-acyclic}&S_{\txt{opt}}=\sum_{\vec{n},\vec{u}_{\hat{n}_1},i} (n_1+1)\pi_i~p_i\left(\mathop{\cap}\limits_{t_1=1}^{n_1}\widehat{R}_{u_{1t_1}=c|\td{u}_{1\td{t}_1}}\right)\\
&~~\times\prod_{k=2}^Kp_i\left(R_{u_{kn_k}|\td{u}_{k\td{n}_k}}\cap~\mathop{\cap}\limits_{t_k=1}^{n_k-1}\widehat{R}_{u_{kt_k}=c|\td{u}_{k\td{t}_k}}\right)\\
&~~\leq \sum_{\vec{n},\vec{u}_{\hat{n}_1},i} (n_1+1)\pi_i~p_i\left(\prod_{t_1=1}^{n_1}R^{\txt{(sbs)}}_{u_{1t_1}=c|\td{u}_{1\td{t}_1}}\right)\\
&~~\times\prod_{k=2}^Kp_i\left(R^{\txt{(sbs)}}_{u_{kn_k}|\td{u}_{k\td{n}_k}}\times~\prod_{t_k=1}^{n_k-1}R^{\txt{(sbs)}}_{u_{kt_k}=c|\td{u}_{k\td{t}_k}}\right)\\
&~~=\sum_{\vec{n},\vec{u}_{\hat{n}_1},i} (n_1+1)\pi_i~\prod_{t_1=1}^{n_1}p_i\left(R^{\txt{(sbs)}}_{u_{1t_1}=c|\td{u}_{1\td{t}_1}}\right)\\
&~~\times\prod_{k=2}^K\left[p_i\left(R^{\txt{(sbs)}}_{u_{kn_k}|\td{u}_{k\td{n}_k}}\right)\prod_{t_k=1}^{n_k-1}p_i\left(R^{\txt{(sbs)}}_{u_{kt_k}=c|\td{u}_{k\td{t}_k}}\right)\right].
\ea\ee
The probability in (\ref{distr-terms-expansion-acyclic-opt}) for the sample-by-sample test is

{\small
\be\ba
\label{distr-terms-expansion-acyclic-opt-sbs}&p^{(\txt{sbs})}_i(u_{1n_1}=j)=\sum_{\hat{n}_1,u_1^{n_1-1},\vec{u}^{\hat{n}_1},\vec{x}^{\vec{n}}} p_i(\vec{x}^{\vec{n}})\\
&~~\times q_{\txt{opt}}(u_{1n_1}=j|x_{1n_1},\td{u}_{1\td{n}_1})\prod_{t_1=1}^{n_1-1}q_{\txt{opt}}(u_{1t_1}=c|x_{1t_1},\td{u}_{1\td{t}_1})\\
&~~\times\prod_{k=2}^K\left(q_{\txt{opt}}(u_{kn_k}|x_{kn_k},\td{u}_{k\td{n}_k})\prod_{t_k=1}^{n_k-1}q_{\txt{opt}}(u_{kt_k}=c|x_{kt_k},\td{u}_{k\td{t}_k})\right)\\
&~~=\sum_{\hat{n}_1,\vec{u}_{\hat{n}_1}} p_i\left(R^{\txt{(sbs)}}_{u_{1n_1}=j|\td{u}_{1\td{n}_1}}\right)\prod_{t_1=1}^{n_1-1}p_i\left(R^{\txt{(sbs)}}_{u_{1t_1}=c|\td{u}_{1\td{t}_1}}\right)\\
&~~\times\prod_{k=2}^K\left[p_i\left(R^{\txt{(sbs)}}_{u_{kn_k}|\td{u}_{k\td{n}_k}}\right)\prod_{t_k=1}^{n_k-1}p_i\left(R^{\txt{(sbs)}}_{u_{kt_k}=c|\td{u}_{k\td{t}_k}}\right)\right].
\ea\ee
}

\begin{rmks}~

1) For simplicity, we have assumed throughout that the prior $\pi$ and error levels $\al,\beta$ are independent of $n$. However, in the detection of time-dependent signals in time-dependent noise (such as a time-dependent $\sigma$ in Example \ref{S-test-example}), we may have to choose these parameters to depend on $n$ as with the treatment of \emph{quickest detection} problems in \cite{siryaev-71,teneketzis-varaiya-84}. Moreover, continuous time versions of the discussed problems can also be considered in the same spirit as in \cite{siryaev-71,lavigna-makowski-baras-86}.

2) The tests we have considered are of infinite horizon in the sense that there is no limit on the stopping time. However, their finite horizon versions can be considered on the same footing and with necessary adjustments.
\end{rmks}  

\section{Conclusion}\label{cnl-section}
By studying both centralized and distributed sequential detection with stopping time as the objective function, we have obtained Wald's classical sequential probability ratio test as a byproduct of the optimal decision rule. Based on the possibility of sample-by-sample processing, we derived upper bounds on the optimal stopping time and used these bounds to study the qualitative dependence of stopping time on observation quality and desired accuracy level.

The distributed sequential test derived for a two-sensor tandem network generalizes in a straightforward way to any distributed sensor network in the form of an acyclic directed graph. This extension was deduced with the help of the usual binary decision rules for static detection over such sensor networks.

\appendices
\section{Proof of Theorem \ref{centr-thm}}\label{centr-thm-proof}
By (\ref{optimal-rule2}) and (\ref{optimal-region2}) and the discussion at the beginning of Section \ref{dp-section}, the decision rule has the form (\ref{centr-dec-rule}), where the decision regions are obtained as follows. For the Lagrangian $L$ in (\ref{centr-lagrangian}), consider the derivatives
\be\ba
\label{centr-proof-der}&{1\over n+1}{\del L\over\del q(u_n=0|x^n)}=-\ld_{n1}p_1(x^n)Q_n(x^{n-1}),\\
&{1\over n+1}{\del L\over\del q(u_n=c|x^n)}\sr{(s)}{=}\big[p_0(x^n)\td{\pi}_{0,n}+p_1(x^n)\td{\pi}_{1,n}\big]\\
&~~~~\times Q_n(x^{n-1}),\\
&{1\over n+1}{\del L\over\del q(u_n=1|x^n)}=-\ld_{n0}p_0(x^n)Q_n(x^{n-1}),
\ea\ee
where $\td{\pi}_{i,n}=(1-\ld)c_{i,n}\pi_i$ for constants $c_{i,n}$, and $Q_n(x^{n-1})=\prod_{t=1}^{n-1}q(u_t=c|x^t)$. Note that step (s) in (\ref{centr-proof-der}) requires the conditional independence assumption in the theorem statement. In terms of these derivatives, the decision regions are
\be\ba
& R_{u_n=0}=\left\{x^n:{\del L\over\del q(u_n=0|x^n)}<{\del L\over\del q(u_n=c|x^n)}\right\}\nn\\
&~~\cap\left\{x^n:{\del L\over\del q(u_n=0|x^n)}<{\del L\over\del q(u_n=1|x^n)}\right\}\nn\\
& R_{u_n=c}=\left\{x^n:{\del L\over\del q(u_n=c|x^n)}<{\del L\over\del q(u_n=0|x^n)}\right\}\nn\\
&~~\cap\left\{x^n:{\del L\over\del q(u_n=c|x^n)}<{\del L\over\del q(u_n=1|x^n)}\right\}\nn\\
& R_{u_n=1}=\left\{x^n:{\del L\over\del q(u_n=1|x^n)}<{\del L\over\del q(u_n=0|x^n)}\right\}\nn\\
&~~\cap\left\{x^n:{\del L\over\del q(u_n=1|x^n)}<{\del L\over\del q(u_n=c|x^n)}\right\}\nn
\ea\ee
If we choose $\ld,\td{\pi}_{i,n},\ld_{ni}$ such that
\be\ba
\ld<1,~~\ld_{ni}<-\td{\pi}_{i,n},~~~~i=0,1,\nn
\ea\ee
then
\be\ba
& R_{u_n=0}=\left\{x^n:{p_1(x^n)\over p_0(x^n)}<\min\left\{{\td{\pi}_{0,n}\over |\ld_{n1}|-\td{\pi}_{1,n}},{|\ld_{n0}|\over|\ld_{n1}|}\right\}\right\},\nn\\
& R_{u_n=c}=\left\{x^n:{\td{\pi}_{0,n}\over|\ld_{n1}|-\td{\pi}_{1,n}}<{p_1(x^n)\over p_0(x^n)}<{|\ld_{n0}|-\td{\pi}_{0,n}\over\td{\pi}_{1,n}}\right\},\nn\\
& R_{u_n=1}=\left\{x^n:\max\left\{{|\ld_{n0}|\over|\ld_{n1}|},{|\ld_{n0}|-\td{\pi}_{0,n}\over\td{\pi}_{1,n}}\right\}<{p_1(x^n)\over p_0(x^n)}\right\}.\nn
\ea\ee
The constraint $R_{u_n=0}\cup R_{u_n=c}\cup R_{u_n=1}=\X^n$ requires that
\be\ba
\label{c-thresholds}&\min\left\{{\td{\pi}_{0,n}\over |\ld_{n1}|-\td{\pi}_{1,n}},{|\ld_{n0}|\over|\ld_{n1}|}\right\}={\td{\pi}_{0,n}\over|\ld_{n1}|-\td{\pi}_{1,n}}\triangleq A_n,\\
& {|\ld_{n0}|-\td{\pi}_{0,n}\over\td{\pi}_{1,n}}=\max\left\{{|\ld_{n0}|\over|\ld_{n1}|},{|\ld_{n0}|-\td{\pi}_{0,n}\over\td{\pi}_{1,n}}\right\}\triangleq B_n.
\ea\ee
Thus, along with the constraints
\be\ba
\label{c-threshold-constraints}&{\td{\pi}_{0,n}\over|\ld_{n1}|-\td{\pi}_{1,n}}\leq{|\ld_{n0}|\over|\ld_{n1}|}\leq {|\ld_{n0}|-\td{\pi}_{0,n}\over\td{\pi}_{1,n}},\\
&\ld<1,~~\ld_{n1}<-\td{\pi}_{1,n},~~\ld_{n0}<-\td{\pi}_{0,n},
\ea\ee
the decision regions take the final form (\ref{centr-dec-regions}), where the thresholds $A_n,B_n$ are given by (\ref{c-thresholds}).

\section{Proof of Theorem \ref{distr-thm}}\label{distr-thm-proof}
By (\ref{optimal-rule2}) and (\ref{optimal-region2}) and the discussion at the beginning of Section \ref{dp-section}, the decision rule at X has the form (\ref{distr-dec-rule-X}), and the decision rule at Y has the form (\ref{distr-dec-rule-Y}).

To determine the decision regions at $Y$, we first evaluate the following derivatives of the probabilities in (\ref{distr-terms-expansion}). With the conditional independence assumptions, we have
\be\ba
& {\del p(v_m=k|H_i)\over\del q(v_m=k'|y^m,u_n)}\sr{CIO}{=}\delta_{kk'}Q_m(y^{m-1},u_n)~\big(c_{i,m}\big)^{\delta_{k'c}}\\
&~~~~\times \sum_{x^n}p(u_n=k'|x^n)p_i(x^n,y^m),\\
& {\del p(v_m=k|H_i)\over\del q(u_n=k'|x^n)}\sr{CIO}{=}Q_n(x^{n-1})~\big(c_{i,n}\big)^{\delta_{k'c}}\\
&~~~~\times\sum_{y^m} p(v_m=k|y^m,u_n=k')p_i(x^n,y^m),
\ea\ee
where $Q_m(y^{m-1},u_n)=\prod_{s=1}^{m-1}q(v_s=c|y^s,u_n)$, $Q_n(x^{n-1})=\prod_{t=1}^{n-1}q(u_t=c|x^t)$, $c_{i,m},c_{i,n}$ are constants, and \emph{CIO} denotes ``\emph{conditionally independent observations}'' (needed for the case where $k'=c$). Using these, we obtain derivatives of the Lagrangian $L$ in (\ref{distr-lagrangian}) as

{\small
\be\ba
&{1\over m+1}{\del L\over\del q(v_m=0|y^m,u_n)}=-\ld_{m1}\sum_{x^n}p(u_n|x^n)p_1(x^n,y^m)\nn\\
&~~~~\times Q_m(y^{m-1},u_n)\\
&~~\sr{CIO}{=}-\ld_{m1}p_1(u_n)p_1(y^m)~Q_m(y^{m-1},u_n),\nn\\
&{1\over m+1}{\del L\over\del q(v_m=c|y^m,u_n)}\sr{CIO}{=}\sum_{x^n}\bigg[\td{\pi}_{0,m}p(u_n|x^n)p_0(x^n,y^m)\nn\\
&~~~~+\td{\pi}_{1,m}p(u_n|x^n)p_1(x^n,y^m)\bigg]Q_m(y^{m-1},u_n)\nn\\
&~~\sr{CIO}{=}\big[\td{\pi}_{0,m}p_0(u_n)p_0(y^m)+\td{\pi}_{1,m}p_1(u_n)p_1(y^m)\big]Q_m(y^{m-1},u_n),\nn\\
&{1\over m+1}{\del L\over\del q(v_m=1|y^m,u_n)}=-\ld_{m0}\sum_{x^n}p(u_n|x^n)p_0(x^n,y^m)\nn\\
&~~~~\times Q_m(y^{m-1},u_n)\sr{CIO}{=}-\ld_{m0}p_0(u_n)p_0(y^m)~Q_m(y^{m-1},u_n),\nn
\ea\ee
}where $\td{\pi}_{i,m}=(1-\ld)c'_{i,m}\pi_i$ for constants $c'_{i,m}$, and $p_i(u_n)=\sum_{x^n}p(u_n|x^n)p_i(x^n)$. In terms of the above derivatives, the decision regions at $Y$ are

{\footnotesize
\be
R_{v_m=k|u_n}\!=\!\bigcap_{k'\neq k}\!\left\{y^m:{\del L\over\del q(v_m=k|y^m,u_n)}\!<\!{\del L\over\del q(v_m=k'|y^m,u_n)}\right\}\nn
\ee
}After some simplification and imposing constraints similar to (\ref{c-threshold-constraints}), the regions take the form in (\ref{distr-dec-regions-Y}), where the thresholds therein are given by
\be
\label{d-thresholds-Y}A^{(2)}_{u_n}={\td{\pi}_{0,m}\over|\ld_{m1}|-\td{\pi}_{1,m}}~{p_0(u_n)\over p_1(u_n)},~~B^{(2)}_{u_n}={|\ld_{m0}|-\td{\pi}_{0,m}\over\td{\pi}_{1,m}}~{p_0(u_n)\over p_1(u_n)}.
\ee

To determine the decision regions at $X$, we likewise evaluate the following derivatives.
\be\ba
&{1\over Q_n(x^{n-1})}{\del L\over\del q(u_n|x^n)}\nn\\
&~~\sr{CIO}{=}\sum_{m,y^m,i}(m+1)p(v_m=c|y^m,u_n)p_i(x^n,y^m)\hat{\pi}_{i,n}\nn\\
&~~-\sum_{m,y^m}\ld_{m0}(m+1)p(v_m=1|y^m,u_n)p_0(x^n,y^m)\nn\\
&~~-\sum_{m,y^m}\ld_{m1}(m+1)p(v_m=0|y^m,u_n)p_1(x^n,y^m)\nn\\
&~\!\sr{CIO}{=}\!\sum_i\Pi_i(u_n)p_i(x^n)-\Ld_0(u_n)p_0(x^n)-\Ld_1(u_n)p_1(x^n)\nn\\
&~=[\Pi_0(u_n)-\Ld_0(u_n)]p_0(x^n)+[\Pi_1(u_n)-\Ld_1(u_n)]p_1(x^n)\nn\\
&~=C_{0u_n}p_0(x^n)+C_{1u_n}p_1(x^n),\nn
\ea\ee
where
\be\ba
&\Pi_i(u_n)=\sum_{m,y^m}(m+1)p(v_m=c|y^m,u_n)p_i(y^m)\hat{\pi}_{i,n},\\
&\Ld_0(u_n)=\sum_{m,y^m}\ld_{m0}(m+1)p(v_m=1|y^m,u_n)p_0(y^m),\\
&\Ld_1(u_n)=\sum_{m,y^m}\ld_{m1}(m+1)p(v_m=0|y^m,u_n)p_1(y^m),\\
& C_{iu_n}=\Pi_i(u_n)-\Ld_i(u_n),\nn
\ea\ee
i.e.,
\be\ba
&{1\over Q_n(x^{n-1})}{\del L\over\del q(u_n=0|x^n)}=C_{00}p_0(x^n)+C_{10}p_1(x^n),\nn\\
&{1\over Q_n(x^{n-1})}{\del L\over\del q(u_n=c|x^n)}\sr{CIO}{=}C_{0c}p_0(x^n)+C_{1c}p_1(x^n),\nn\\
&{1\over Q_n(x^{n-1})}{\del L\over\del q(u_n=1|x^n)}=C_{01}p_0(x^n)+C_{11}p_1(x^n),\nn
\ea\ee
where $\hat{\pi}_{i,n}=(1-\ld)c'_{i,n}\pi_i$ for constants $c'_{i,n}$. In terms of these derivatives, the decision regions at $X$ are

{\small
\bea
R_{u_n=k}=\bigcap_{k'\neq k}\left\{x^n:{\del L\over\del q(u_n=k|x^n)}<{\del L\over\del q(u_n=k'|x^n)}\right\}.\nn
\eea
}After some simplification and applying the constraints
\be\ba
C_{10}>C_{1c}>C_{11},~~~~C_{01}>C_{0c}>C_{00},~~~~C_{10}=C_{01},\nn
\ea\ee
these regions take the form in (\ref{distr-dec-regions-X}), where the thresholds therein are given by
\be\ba
\label{d-thresholds-X}& A^{(1)}=\min\left\{{C_{0c}-C_{00}\over C_{10}-C_{1c}},{C_{01}-C_{00}\over C_{10}-C_{11}}\right\}={C_{0c}-C_{00}\over C_{10}-C_{1c}},\\
& B^{(1)}={C_{10}-C_{0c}\over C_{1c}-C_{11}}=\max\left\{{C_{01}-C_{0c}\over C_{1c}-C_{11}},{C_{01}-C_{00}\over C_{10}-C_{11}}\right\}.
\ea\ee


%

\end{document}